\documentclass[aps,prb,twocolumn,superscriptaddress]{revtex4-1}
\usepackage[utf8]{inputenc}
\usepackage{amsmath}
\usepackage{braket}
\usepackage{bm}
\usepackage{graphicx}
\usepackage{verbatim}
\usepackage{xcolor}

\usepackage[bookmarks=true,colorlinks=true,citecolor=blue,urlcolor=blue]{hyperref}

\usepackage[percent]{overpic}

\renewcommand{\Im}{\operatorname{Im}}

\newcommand{\be}{\begin{equation}}
\newcommand{\ee}{\end{equation}}
\newcommand{\bea}{\begin{eqnarray}}
\newcommand{\eea}{\end{eqnarray}}

\begin{document}
\title{Hydrodynamics of interacting spinons in the magnetized spin-$1/2$ chain with the uniform Dzyaloshinskii-Moriya interaction}

\author{Ren-Bo Wang}
\affiliation{Department of Physics and Astronomy, University of Utah,   Salt Lake City, Utah 84112, USA}
\author{Anna Keselman}
\thanks{The Lawrence S. Jackier Fellow}
\affiliation{Physics Department, Technion, 32000 Haifa, Israel}
\author{Oleg A. Starykh}
\affiliation{Department of Physics and Astronomy, University of Utah,   Salt Lake City, Utah 84112, USA}

\date{\today}

\begin{abstract}
We use a hydrodynamic approach to investigate dynamic spin susceptibility of the antiferromagnetic spin-$1/2$ Heisenberg chain with a uniform Dzyaloshinskii-Moriya (DM) interaction in the presence of an external magnetic field. We find that transverse (with respect to the magnetic field) spin susceptibility harbors two (respectively, three) spin excitation modes when the magnetic field is parallel (respectively, orthogonal) to the DM axis. In all cases, the marginally irrelevant backscattering interaction between the spinons creates a finite energy splitting between optical branches of excitations at $k=0$. Additionally, for the orthogonal geometry, the two lower spin branches exhibit avoided crossing at finite momentum which is determined by the total magnetic field  (the sum of the external and internal molecular fields) acting on spinons. Our approximate analytical calculations compare well with numerical results obtained using matrix-product-state (MPS) techniques. Physical consequences of our findings for the electron spin resonance experiments are discussed in detail.
\end{abstract}

\maketitle

\section{Introduction}
\label{sec:intro}
Quantum spin liquids (QSL) continue to attract widespread interests of physicists  
due to numerous novel features arising from their topological characters such as long-ranged quantum entanglements, fractional exciations, and emergent gauge fields \cite{Lee2007,Savary2016,Knolle2019,Broholm2020} as well as promising application to topological quantum computations \cite{Nayak2008,Lahtinen2017}. 
The antiferromagnetic spin-$1/2$  chain \cite{Bethe1931} with its critical ground state without conventional long-range magnetic order but with long-range (power-law) correlations  serves as a paradigmatic model of a QSL in one-dimension ($1d$).
The elementary excitations of the spin chain, neutral spinons with spin-$1/2$,  exhibit two-spinon continuum which have been observed in inelastic neutron scattering measurements of various quasi-$1d$ spin-$1/2$ antiferromagnets such as, for example, CuSO$_ 4\cdot$5D$_ 2$O  \cite{Mourigal2013} and KCuF$_ 3$ \cite{Lake2013}. 
Unexpected doublet-like structure of the spinon continuum near zero momentum, discovered in electron spin resonance (ESR) experiments  \cite{Povarov2011,Smirnov2015}, was explained by the internal spin-orbital field produced by the uniform Dzyaloshinskii-Moriya (DM) interaction \cite{Dzyaloshinsky1958,Moriya1960}. 

More recently, some of the previously unexplained features of the small-momentum spinon response, such as  a field-dependent finite energy splitting of the spinon continuum at $k=0$  and the curved dispersions of the spin-1 excitations at small $k$, noticed both experimentally \cite{Stone2003} and numerically \cite{Kohno2009}, were explained as originating from the backscattering interaction $g_\mathrm{bs}$ between spinons in finite magnetic field \cite{Keselman2020}.

In this manuscript, we develop this point of view further by re-formulating it as hydrodynamics of magnetization densities and currents. 
This hydrodynamic formulation provides for a very efficient description  of the dynamical susceptibility of the spin chain with the uniform DM interaction and subject to the external magnetic field oriented at an arbitrary angle with the DM axis. 
We show that the inter-spinon interaction produces qualitative changes to the non-interacting spinon picture \cite{Povarov2011} and describe its key consequences for ESR experiments.

The paper is organized as follows. 
Sections \ref{sec:1} and \ref{sec:2} describe the spin model and its low-energy field-theoretic formulation in terms of chiral spin currents. 
Section \ref{sec:hydro} explains the hydrodynamic approximation that is used in Sec.\ \ref{sec:GF} to derive dynamic spin susceptibility $\chi(k,\omega)$ at small momenta for the important cases of the parallel (${\bf h} \parallel  {\bf D}$) and orthogonal (${\bf h} \perp {\bf D}$) orientations between the magnetic field and the DM axis. 
For the arbitrary angle between them we, for simplicity, restrict the consideration to $\chi(k=0,\omega)$.
Physical consequences of the backscattering interaction for ESR experiments are described in Sec.\ \ref{sec:esr}. 
Our analytical results are critically compared with 
accurate, unbiased numerical results obtained using matrix-product-state (MPS) techniques
in Section \ref{sec:dmrg}. 
Section \ref{sec:conclude} concludes the manuscript. 
Some of the more technical results are presented in Appendices.

Throughout the paper operators are denoted by hats on top of them, vectors are denoted by bold letters, and calligraphic letters are reserved for matrices.

\section{The Model}
\label{sec:1}
Consider a $1d$ antiferromagnetic spin-$1/2$ Heisenberg chain with a  uniform DM interaction $\mathbf{D}$ in the presence of an  external magnetic field ${\bf H}$ \cite{Garate2010,Povarov2011,Karimi2011,Chan2017}
\be
\hat H
=\sum_n \Big(J\hat {\mathbf{S}}_n\cdot\hat{\mathbf{S}} _{n+1}
-\mathbf D\cdot \hat{\bf S}_n \times \hat {\bf S}_{n+1}
-\mathbf{h}\cdot\hat{ \mathbf{S}}_n\Big),\label{2.1}
\ee
where $\mathbf{h}=g\mu_B\mathbf{H}$.

In the case of the magnetic field parallel to the DM axis, say, the $z$-axis, an important general consideration is possible on the level of the lattice Hamiltonian. 
We carry out unitary transformation to rotate spins about $z$-axis as \cite{Povarov2011,Aristov2000,Bocquet2001}
\begin{align}
\hat S^+_n 
=\hat{ \tilde{S}}^+_n e^{-i k_{\rm dm} n a}, \quad
\hat S^z_n 
= \hat {\tilde{S}}^z_n ,\label{2.2}
\end{align}
where $a$ is the lattice constant and
\begin{align}
k_{\rm dm} 
= \tan^{-1}(D/J)/a 
\approx {D}/({J a}).\label{2.3}
\end{align}
In the following, we set $a=1$. 
The  Hamiltonian \eqref{2.1} transforms into 
\begin{align}
\begin{aligned}
\hat{\tilde{H}}
=&\tilde{J}\sum_n[\frac{1}{2}(\hat{\tilde{S}}^+_{n}\hat{\tilde{ S}}^-_{n+1}+\hat{\tilde{S}}^-_{n}\hat{\tilde{S}}^+_{n+1})
+\Delta\hat{\tilde{S}}_{n}^z\hat{\tilde{S}}_{n+1}^z]-h\sum_n\hat{ \tilde{S}}_{n}^z.
\end{aligned}\label{2.4}
\end{align}
Here we see that \eqref{2.4} is just a chain without the DM interaction with exchange interaction 
$\tilde{J}=\sqrt{J^2+D^2} \approx J + D^2/(2J)$ 
and anisotropy parameter 
$\Delta={J}/\tilde{J} \approx 1 - D^2/(2J^2)$. 
For the chain with $D \ll J$, which is the case of our interest, these quadratic deviations can be neglected.

The most important consequence of the simple transformation \eqref{2.2} is that the dynamic structure factor ${\cal S}^{+-}(k, \omega)$ of the Hamiltonian \eqref{2.1},
\be
{\cal S}^{+-}(k, \omega) = \sum_n \int dt e^{i \omega t} e^{-i k n a} \langle \hat{S}^+_n(t) \hat{S}^-_0(0)\rangle_{\hat{H}} ,
\label{o2}
\ee
where the expectation value $\langle ... \rangle_{\hat{H}}$ is taken with the respect to the equilibrium density matrix of the Hamiltonian $\hat{H}$ \eqref{2.1},
reduces to that  of the rotated $\hat{\tilde{H}}$ \eqref{2.4},
%$\tilde{{\cal S}}^{+-}(q + q_{\rm dm}, \omega)$, 
\bea
\tilde{{\cal S}}^{+-}(k+k_{\rm dm}, \omega) &=& \sum_n \int dt e^{i \omega t} e^{-i k n a} \nonumber\\
&&\times \langle e^{-i k_{\rm dm} n a} \hat{\tilde{S}}^+_n(t) \hat{\tilde{S}}^-_0(0)\rangle_{\hat{\tilde{H}}} ,
\label{o3}
\eea
but with the {\em boosted} momentum $k + k_{\rm dm}$. 

The same relation also apply to the transverse dynamical susceptibility, defined by the retarded Green's function of the spin operators $S^+_n$ and $S^-_0$,
\be
\chi^{+-}(k,\omega) = -i \sum_n \int_0^\infty dt e^{i \omega t} e^{-i k n a} \langle [ \hat S^+_n(t),\hat S^-_0(0)] \rangle_{\hat{H}}  .
\label{o5}
\ee
It is connected with the dynamic structure factor by the Fluctuation Dissipation theorem,
\be
{\cal S}^{+-}(k, \omega) = -2 (n(\omega) + 1) {\rm Im}[\chi^{+-}(k,\omega)].
\label{ou6}
\ee
Here $n(\omega) = 1/(e^{\omega/T} -1)$ is the Bose function so that in the zero-temperature limit, $T\to 0$, the right-hand-side of \eqref{ou6} is non-zero only for $\omega > 0$.

The equivalence of the structure factors \eqref{o2} and \eqref{o3} translates into that of the susceptibilities \cite{Povarov2011,Karimi2011},
\be
\chi^{+-}(k,\omega) = \tilde{\chi}^{+-}(k+k_{\rm dm},\omega),
\label{op7}
\ee
where $\tilde{\chi}^{+-}(q,\omega)$ is the transverse susceptibility of the chain described by $\hat{\tilde{H}}$ \eqref{2.4} (equivalently, within our approximation of neglecting $D^2/J^2 \to 0$ in \eqref{2.4}, by Eq.\ \eqref{2.1} with {\em no} DM term, $D=0$).

It is also easy to see that for the transverse susceptibility for the opposite, ``$-+$", circulation, 
\be
\chi^{-+}(k,\omega) = -i \sum_n \int_0^\infty dt e^{i \omega t} e^{-i k n a} \langle [\hat S^-_n(t),\hat S^+_0(0)] \rangle_{\hat{H}} ,
\label{o8}
\ee
the DM-induced shift occurs in the {\em opposite} direction,
\be
\chi^{-+}(k,\omega) 
= \tilde{\chi}^{-+}(k - k_{\rm dm},\omega).
\label{o9}
\ee
Finally, the longitudinal susceptibility does not experience the DM-induced shift of $k$ at all, $\chi^{zz}(k,\omega) = \tilde{\chi}^{zz}(k,\omega)$.

This crucial feature of the spin chain with the uniform DM interaction turns the standard ESR experiment, which measures $k=0$ response, into a finite-momentum probe of the dynamic correlations at $k=k_{\rm dm}$  and allows us to explore details of the small-momentum response of the spin-1/2 chain in the magnetic field with accuracy greatly exceeding that of the inelastic neutron scattering experiments.

\section{Low-energy description}
\label{sec:2}

Within the field-theoretic description of the spin chain spin operators are approximated by the sum of uniform and staggered components \cite{Garate2010,Chan2017}
\begin{align}
\hat{\bf S}_n
\to a[\hat{\bf J}_L(x)+\hat{\bf J}_R (x) + (-1)^{x/a} \hat{\bf N}(x)],
\label{os40}
\end{align}
where $x=n a$ is the coordinate of the $n$th spin along the chain,
$\hat{\bf J }_{R/L}$ is the right/left ($R/L$) chiral spin current, describing the uniform spin density, and $\hat{\bf N}$ is the staggered (N\'eel) component of the spin density.
Spin currents obey the Kac-Moody algebra  \cite{Ludwig1995,Balents2001}
\bea
[\hat J_{R/L}^a(x),\hat J_{R/L}^b(x')] &=& \frac{\mp i}{4\pi} \delta'(x-x')\delta^{ab} +\nonumber\\
&&+ i \delta(x-x')  \epsilon^{abc}\hat J_{R/L}^c(x) ,
\label{eq1}
\eea
where prime on the delta function denotes derivative with respect to its argument. Commutation relation \eqref{eq1} is the crucial element of our theory.

The low-energy Hamiltonian of the spin chain \eqref{2.1} is written in the Sugawara form\cite{Gogolin1998}
\begin{align}
\hat H
=&\hat H_0+\hat H_\mathrm{ bs}+\hat V,\label{3.1}\\
\hat H_0
=&\frac{2\pi v}{3}\int\mathrm{d}x \, : \hat{\bf J}_R\cdot \hat{\bf J}_R+\hat{\bf J}_L\cdot {\bf J}_L : ,\label{3.2}\\
\hat{H}_\mathrm{ bs}
=& -g_\mathrm{bs}\int \mathrm{d}x \,:\hat{\bf J}_R\cdot \hat{\bf J}_L : , \label{3.3}\\
\hat{V}
=& - \int \mathrm dx \,\Big({\bf h}\cdot (\hat{\bf J}_R + \hat{\bf J}_L) + \tilde{\mathbf{D}}\cdot (\hat{\bf J}_R - \hat{\bf J}_L)\Big),
%-\sum_{r=R,L}({\bf h}+r\tilde{\mathbf{D}})\cdot\int \mathrm dx\hat{\bf J}_r,
\label{3.4}
\end{align}
where $v = \pi J a/2$ is the spinon velocity and columns $:\,\, :$ denote normal ordering. 
The backscattering interaction, parameterized by the coupling constant $g_\mathrm{bs}$, plays the key role in our study. It describes marginally-irrelevant, in the renormalization group
sense, residual interaction between otherwise independent right- and left- spin currents.
The right-hand-side of \eqref{3.3} is allowed to have an additional term\cite{Garate2010} $\propto \lambda :\hat{J}^d_R \hat{J}^d_L :$, where $\hat{J}^d_{R/L}$ denotes along-the-DM component of the chiral current
 and $\lambda \sim D^2/J^2 \ll 1$ is the anisotropy parameter.
In the case of the weak DM interaction $D \ll J$, which is the natural limit we focus on, this small DM-induced anisotropy can be neglected, $\lambda \to 0$.
This is equivalent to the neglect of $D^2/J^2$ terms in \eqref{2.4}.

The last term, $\hat{V}$ in \eqref{3.4}, describes Zeeman magnetic field ${\bf h}$ and
DM interaction $\tilde{\mathbf{D}}$ acting on spin currents. 
The vector $\tilde{\mathbf{D}}$ is directly proportional to the DM one, $\mathbf{D}$, and the proportionality constant is fixed below.
Notice that the two terms of $\hat{V}$ transform oppositely under the parity $x\to -x$ transformation: the Zeeman term is even under it while the DM term is odd,
in agreement with the lattice Hamiltonian \eqref{2.1}. 

It is convenient to introduce the {\em magnetization} $\hat{\mathbf M}$  and 
the {\em magnetization current} $\hat{\mathbf J}$ operators
\be
\hat{\mathbf M} = \hat {\mathbf{J}} _R +\hat{ \mathbf{J}} _L, \quad
\hat{\mathbf J} = \hat {\mathbf{J}} _R -\hat{ \mathbf{J}} _L
\label{o7}
\ee
in terms of which \eqref{3.4} is expressed as 
\begin{align}
\hat{V} 
=- \int \mathrm dx \, (\mathbf{h} \cdot\hat{\mathbf{M}} 
+\tilde{\mathbf{D}} \cdot \hat{\mathbf{J}} ).
\label{3.8}
\end{align}

\section{Hydrodynamic equations}
\label{sec:hydro} 

Given the commutator \eqref{eq1} and the Hamiltonian \eqref{3.1}, it is easy to write down Heisenberg equations of motion for the chiral spin currents 
$\hat{\bf J}_{R/L}(x,t) = e^{i\hat H t} \hat{\bf J}_{R/L}(x) e^{-i\hat H t}$ (see Appendix \ref{app:A}). 
We find
\bea
&&\partial_t \hat{\bf J}_{R/L}(x,t) = \mp v \partial_x\hat{\bf J}_{R/L}(x,t) - ({\bf h} \pm \tilde{{\bf D}})\times \hat{ \bf J}_{R/L}(x,t)\nonumber\\
&& \pm g_\mathrm{bs}\big(\frac{1}{4\pi}\partial_x\hat{\bf J}_{L/R}(x,t)+\hat{ \bf J}_R(x,t)\times \hat {\bf J}_L(x,t)\big),
\label{4.1}
\eea
where the upper/lower signs apply to right/left currents, correspondingly. The second line of this equation is due to the backscattering interaction \eqref{3.3} between chiral currents.

Taking the sum and the difference of \eqref{4.1}, we readily find equations of motion for the magnetization $\hat{\mathbf M}(x,t)$ and the magnetization current $\hat{\mathbf J}(x,t)$,
\begin{align}
\partial_t\hat{\bf M}(x,t)
=& - v(1+\delta) \partial_x \hat{\bf J}(x,t)\nonumber\\
&-{\bf h}\times\hat{\bf M}(x,t)
-\tilde{\bf D}\times \hat {\bf J}(x,t),
\label{4.2}\\
\partial_t\hat{\bf J}(x,t)
=& - v(1-\delta) \partial_x \hat{\bf M}(x,t)\nonumber\\
&-{\bf h}\times\hat{\bf J}(x,t)
-\tilde{\bf D}\times \hat{ \bf M}(x,t)\nonumber\\
&- 4\pi v \delta \, \hat {\bf M}(x,t)\times\hat{\bf J}(x,t) .
\label{4.3}
\end{align}
Here we introduced dimensionless interaction parameter $\delta = g_\mathrm{bs}/(4\pi v)$. Interaction enters these equations in two different ways. It renormalizes terms with 
spatial derivatives, thanks to the $\partial_x \delta(x-x')$ term in \eqref{eq1}. It also makes equation for the current $\hat{\mathbf J}$ non-linear, as the last line of \eqref{4.3} shows.

It is worth noting that \eqref{4.2} represents the spin continuity equation. Naturally, finite ${\bf h}$ and $\tilde{\mathbf{D}}$ violate the continuity and cause precessional motion of the spin density. They play the role, correspondingly, of the 
temporal and spatial components of the effective background non-Abelian field \cite{Chandra1990,Tokatly2008,Tchernyshyov2021}. 
Eq.\ \eqref{4.2} for the $a$-th component of magnetization $\hat{M}^a$ shows that 
the spatial derivative and the DM field appear in the combination $\partial_x J^a + (\tilde{\bf D}\times \hat {\bf J})^a /(v(1+\delta))$ that is independent of the angle between the
magnetic field ${\bf h}$ and the DM interaction $\tilde{\mathbf{D}}$. This observation, when applied to the case of their parallel orientation ${\bf h} \parallel {\bf D}$, allows
one to fix the coefficient of proportionality between ${\bf D}$ and $\tilde{\mathbf{D}}$, see \eqref{5.8} below.

The Zeeman  and DM fields  \eqref{3.8} induce nonzero equilibrium values of the magnetization and spin current in the ground state. In the non-interacting chain with $g_\mathrm{bs}=0$ they are given by ${\bf m}_0 = \braket{\hat{\bf M}} = \chi_0 {\bf h}$ and ${\bf j}_0 = \braket{\hat{\bf J}} = \chi_0 \tilde{{\bf D}}$, where $\chi_0=1/(2\pi v)$ is the susceptibility of one-dimensional non-interacting Dirac fermions. Due to the opposite parity of the Zeeman and DM terms the two expectation values do not mix with each other.

Finite backscattering interaction induces corrections to these results via internal exchange or ``molecular" fields $\propto g_\mathrm{bs} \braket{\hat{\bf J}_{R/L}}$ acting on L/R currents correspondingly. (The terminology follows Leggett's paper\cite{Leggett1970}.) Within this simple mean-field approximation, we approximate the backscattering \eqref{3.3} as
\begin{align}
\hat H_\mathrm{ bs}
&\approx-g_\mathrm{ bs}\int \mathrm dx({\bf j}_R\cdot \hat {\bf J}_L+\hat{ \bf J}_R\cdot{ \bf j}_L),
\label{eq2}
\end{align}
where ${\bf j}_{R/L} =\braket{\hat {\bf J}_{R/L}}$ is the equilibrium value of the chiral current in the ground state. 
(In the more technical terms, this corresponds to the normal ordering of $\hat H_\mathrm{ bs}$ with respect to the ground state with finite ${\bf j}_{R/L}$.
Diagrammatically, these averages correspond to a tadpole diagram for the fermion self-energy.)
As a result, the effective one-body potential experienced by the currents becomes 
\begin{align*}
\hat V = & -\int \mathrm dx  \,
({\bf h} +\tilde{\bf D} + g_\mathrm{ bs}{\bf j}_L)\hat {\bf J}_R 
+({\bf h } - \tilde{\bf D} + g_\mathrm{ bs}{\bf j}_R) \hat {\bf J}_L.
\end{align*}
This leads to simple self-consistent equations 
%Since the chiral currents respond to the effective fields,
\begin{align*}
\frac{1}{2}\chi_0({\bf h} + \tilde{\bf D} + g_\mathrm{ bs}{\bf j}_L)
&= {\bf j}_R,\\
\frac{1}{2}\chi_0({\bf h} - \tilde{\bf D} + g_\mathrm{ bs}{\bf j}_R)
&={\bf j}_L,
\end{align*}
%where $\chi_0=1/(2\pi v)$ is the non-interacting susceptibility, we find
that are solved by 
\begin{align*}
{\bf j}_{R/L}
=\frac{1}{2}\chi_0
(\frac{\bf h}{1-\delta} \pm \frac{\tilde{\bf D}}{1+\delta}).
\end{align*}
Therefore the equilibrium magnetization ${\bf m}$ of the interacting spinon liquid is
\begin{align}
{\bf m}
&={\bf j}_R+{\bf j}_L = \braket{\hat{\bm M}}
=\frac{\chi_0}{1-\delta}{\bf h},\label{4.5}
\end{align}
while its equilibrium magnetizaton current is 
\begin{align}
{\bf j}
&={\bf j}_R-{\bf j}_L = \braket{\hat{\bm J}}
=\frac{\chi_0}{1+\delta}\tilde{\bf D}, 
\label{4.6}
\end{align}
where $\delta=g_\mathrm{bs}\chi_0/2$ as defined previously.

Equations  of motion \eqref{4.2} and \eqref{4.3} can now be {\em linearized} to the first order in fluctuating quantum fields 
\be
\delta \hat {\bf m}(x,t) \equiv \hat{\bf M}(x,t)-{\bf m}, \quad
\delta \hat {\bf j}(x,t) \equiv\hat{\bf J}(x,t)-{\bf j}.
\label{o5}
\ee
We obtain the following linear vector equations 
\begin{align}
\partial_t\delta \hat {\bf m}(x,t)
=&-v(1+\delta)\partial_x\delta \hat {\bf j}(x,t)\nonumber\\
&-{\bf h}\times \delta \hat {\bf m}(x,t)
-\tilde{\bf D}\times \delta \hat {\bf j}(x,t),\label{4.7}\\
\partial_t\delta \hat {\bf j}(x,t)
=&-v(1-\delta)\partial_x\delta \hat {\bf m}(x,t)\nonumber\\
&-\frac{1-\delta}{1+\delta}\tilde{\bf D}\times\delta \hat {\bf m}(x,t)
-\frac{1+\delta}{1-\delta}{\bf h}\times\delta \hat {\bf j}(x,t),\label{4.8}
\end{align}
where in the last equation we omitted the term $\delta \hat {\bf m}(x,t)\times\delta \hat {\bf j}(x,t)$ as being of the higher (second) order in fluctuations. Note that constant terms 
appearing in the equation for $\delta \hat {\bf j}$ add up to zero, ${\bf h}\times {\bf j} + \tilde{\bf D}\times { \bf m} + 4\pi v \delta {\bf m}\times{\bf j} = 0$, thanks to relations 
\eqref{4.5} and \eqref{4.6}. This constitutes a consistency check of our mean-field approximation \eqref{eq2}. Last two terms in \eqref{4.8} account for ``molecular" field corrections to the DM and Zeeman
interactions, respectively. 
This is easy to see by noting that, for example, $g_\mathrm{ bs} {\bf m} = 2\delta \, {\bf h}/(1-\delta)$ and the fact that $(1+\delta)/(1-\delta) = 1 + 2\delta/(1-\delta)$.

In Fourier space, the linearized hydrodynamic equations \eqref{4.7} and \eqref{4.8} can be written in a compact matrix form
\begin{align}
\omega{\bm\delta}{\hat{ \bm\psi}}(k,\omega)
={\cal A}(k){\bm\delta}{\hat{ \bm\psi}}(k,\omega),\label{4.9}
\end{align}
where we introduce the vector $\bm\delta\hat {\bm \psi}
=
(
\delta \hat m^+,
\delta \hat m^-,
\delta \hat m^z,
\delta \hat j^+,
\delta \hat j^-,
\delta \hat j^z)^\mathrm{T}
$ and a $6\times 6$ matrix
\be
{\cal A}
=\begin{pmatrix}
{\cal A}_h&{\cal A}_{D}\\
\frac{1-\delta}{1+\delta}{\cal A}_{D}&\frac{1+\delta}{1-\delta} {\cal A}_h
\end{pmatrix} ,
\ee
that is composed of $3\times 3$ matrices 
\begin{align}
{\cal A}_h
&=
\begin{pmatrix}
h&0&0\\
0&-h&0&\\
0&0&0
\end{pmatrix},\label{4.10}\\
{\cal A}_{D}
&=\begin{pmatrix}
(1+\delta)vk+\tilde{D}^z&0&-\tilde{D}^+\\
0&(1+\delta)vk-\tilde{D}^z&\tilde{D}^-\\
-\frac{1}{2}\tilde{D}^-&\frac{1}{2}\tilde{D}^+&(1+\delta )vk
\end{pmatrix}.\label{4.11}
\end{align}
Here and in the following, the magnetic field direction is chosen along the $\hat{z}$-axis, ${\bf h} = (0,0,h)$, and transverse components of fluctuating fields are assembled into circular $\pm$ polarizations so that 
$\delta \hat m^\pm = \delta \hat m^x \pm i \delta \hat m^y$, and $\delta \hat j^\pm$ as well as $\tilde{D}^\pm$ are defined similarly.

To check the approach, we first consider the case of the ideal spin chain with ${\bf D}=0$. In this limit matrices ${\cal A}_h, {\cal A}_D$ are diagonal and opposite circular components decouple 
from each other, as well as from the longitudinal fluctuations. We obtain, for example,
\bea
&&(\omega - h) \, \delta \hat m^+ = (1+\delta) v k \,\delta \hat j^+, \nonumber\\
&& (\omega - \frac{1+\delta}{1-\delta} h) \, \delta \hat j^+ = (1-\delta) v k \, \delta \hat m^+ .
\label{o6}
\eea
This simple system of equations reproduces complete spin dispersion relations $\omega_\pm(k)$ \eqref{5.6} derived previously in Ref.\ \onlinecite{Keselman2020}, see also Section \ref{sec:h||d} below. Moreover, it shows that at $k=0$ the uniform magnetization precesses at the Zeeman frequency $\omega_{-}(k=0)=h$, in accordance with the Larmor theorem, while the magnetization current precesses at the {\em higher} frequency $\omega_{+}(k=0)= h (1+\delta)/(1-\delta)$. The residue of the magnetization-current mode at $k=0$ is, however, exactly zero, $A_{+}(k=0)=0$ \cite{Keselman2020}, see also \eqref{5.7} in Sec.\ \ref{sec:h||d} below.
At finite $k$ the two modes hybridize. 

\section{Green's functions}
\label{sec:GF}

Physics of the problem is encoded in the dynamical susceptibility which is given by the matrix of retarded Green's functions
\begin{align}
{\cal G}^{ab}(x,t;x',t')
&=-i\theta(t-t')\braket{[\delta \hat \psi^a(x,t),\delta \hat \psi^b(x',t')]}.
\label{os35}
\end{align} 
%where $\hat{ \bm \psi} = (\hat M^+,\hat M^-,\hat M^z,\hat J^+,\hat J^-,\hat J^z)^\mathrm{T}$
It obeys the standard equation of motion 
\begin{align}
\begin{aligned}
&\partial_t {\cal G}^{ab}(x,t;x',t')\\
=&-i\delta(t-t')\braket{[\delta\hat \psi^a(x,t),\delta\hat \psi^b(x',t)]}\\
&-i\theta(t-t')\braket{[\partial_t\delta \hat \psi^a(x,t),\delta \hat \psi^b(x',t')]}.
\end{aligned}\label{5.1}
\end{align}
In Fourier space, Eq.\ \eqref{5.1} is solved with the help of \eqref{4.9} in a compact form
\begin{align}
{\cal G}(k,\omega)
&=[\omega-{\cal A}(k) + i 0^+]^{-1}{\cal F}(k),\label{5.2}
\end{align}
where the matrix of commutators is given by 
\be
{\cal F}
=\begin{pmatrix}
{\cal F}_m&{\cal F}_j\\
{\cal F}_j&{\cal F}_m
\end{pmatrix}
\label{eq3}
\ee
with
\begin{align}
{\cal F}_m
&=\begin{pmatrix}
0
&2m
&0\\
-2m
&0
&0\\
0
&0
&0
\end{pmatrix},\label{27}\\
{\cal F}_j
&=
\begin{pmatrix}
0
&\frac{k}{\pi}+ 2j^z
&-j^+\\
\frac{k}{\pi}-2j^z
&0
&j^-\\
j^+
&-j^-
&\frac{k}{2\pi}
\end{pmatrix}.
\label{o36}
\end{align}

%added May 15, 2022, upon request of the Referee
\subsection{Brief overview}
\label{sec:key}

The retarded Green's function depends strongly on the relative orientations between ${\bf h}$ and ${\bf D}$. 
Below we discuss  transverse susceptibilities $\chi^{+-}(k,\omega)={\cal G}^{12}(k,\omega),\chi^{-+}(k,\omega)={\cal G}^{21}(k,\omega)$ as well as the longitudinal susceptibility 
$\chi^{zz}(k,\omega)={\cal G}^{33}(k,\omega)$ for specific cases ${\bf h}\parallel {\bf D}$ and ${\bf h}\perp{\bf D}$, and then present analytical result for ${\cal G}(k=0,\omega)$ for 
the general case of the arbitrary angle between ${\bf h}$ and ${\bf D}$.

In Section \ref{sec:h||d}, we discuss the parallel geometry, ${\bf h}\parallel {\bf D}$, which is the simplest case. 
In agreement with the unitary transformation argument of Sec.\ \ref{sec:1}, we find below that finite DM simply shifts the wavevector of the transverse susceptibility by $k_{\rm dm}$ but otherwise does not affect the two-mode structure of $\chi^{+-}$.

Tilting ${\bf h}$ away from ${\bf D}$ destroys the $U(1)$ symmetry of the problem and couples magnetization and magnetization-current modes with the longitudinal one, resulting in the three-pole structure of the susceptibility. In the case of the perpendicular geometry, ${\bf h}\perp{\bf D}$, in Sec.\ \ref{sec:DperpH}, the $k$-dependence of these coupled spin modes and their spectral weights can be understood in much details analytically. One of the interesting findings there is the avoided crossing between modes $\omega_1$ and $\omega_2$, which takes place at finite $k$, as illustrated in Fig.\ \ref{fig:2}.

The case of the arbitrary angle between ${\bf h}$ and ${\bf D}$ is presented in Sec.\ \ref{sec:arbDH}. Here, calculations at finite $k$ become too complicated algebraically and we focus on the ESR-related $k=0$ limit only. 
In this limit the susceptibility \eqref{o33} can again be expressed in terms of two modes $\Omega_\pm$ \eqref{5.15} (the third mode, as well as its residue, vanish at $k=0$). 

These findings make it possible to discuss ESR in Sec.\ \ref{sec:esr} and open the way for the direct comparison with the unbiased numerical simulations based on matrix-product-state techniques in Section \ref{sec:dmrg}. 

\subsection{${\bf h}\parallel {\bf D}$}
\label{sec:h||d}

For ${\bf h}\parallel {\bf D}$, we set $\tilde{\bf D}=\tilde{D}\hat{\bf z}$ in \eqref{5.2} and obtain for the transverse susceptibility
\begin{align}
&\chi^{+-}(k,\omega)
=\chi_0\big(\frac{A_+(k)}{\omega-\omega_{+}(k)+i 0^+}
+\frac{A_-(k)}{\omega-\omega_{-}(k)+i 0^+}\big) \label{5.3}\\
&\omega_{\pm}(k)
=\frac{h}{1-\delta} \pm\sqrt{(\frac{\delta h}{1-\delta})^2+(1-\delta^2)v^2\big(k + \frac{\tilde{D}}{v(1+\delta)}\big)^2},\label{5.6}\\
&A_\pm(k)
=\frac{h}{1-\delta}\pm\frac{-\delta(\frac{h}{1-\delta})^2+(1+\delta)v^2\big(k + \frac{\tilde{D}}{v(1+\delta)}\big)^2}{\sqrt{(\frac{\delta h}{1-\delta})^2+(1-\delta^2)v^2\big(k + \frac{\tilde{D}}{v(1+\delta)}\big)^2}}.
\label{5.7}
\end{align}
Observe that $k$ shows up only in the combination $\tilde{k} = k + \tilde{D}/(v(1+\delta))$ in these equations. 
The unitary rotation argument in Section \ref{sec:1} tells us that momentum $k$ is boosted as $k \to \tilde{k} = k + k_{\rm dm}$, see \eqref{2.3}.
%is the boosted momentum, in full agreement with the unitary rotation argument in Section \ref{sec:1}.
This allows us to identify the momentum boost $k_{\rm dm}=D/J $ with $\tilde{D}/(v(1+\delta))$ and thereby obtain 
the relation between the DM parameter of the lattice Hamiltonian \eqref{2.1} and the parameter $\tilde{D}$ of the continuum low-energy theory \eqref{3.4},
\begin{align}
\tilde{D} = v (1+\delta) \frac{D}{J} .\label{5.8}
\end{align}
For sufficiently small magnetic field $v \approx \pi J/2$ and thus $\tilde{D} \approx \pi (1+\delta) D/2$. Transverse spin susceptibility for the opposite circulation, $\chi^{-+}$, follows from the Onsager's relation (time-reversal transformation), $\chi^{-+}(k,\omega)|_h = \chi^{+-}(-k,\omega)|_{-h}$ (do not confuse $k$ with $\tilde{k}$ here).

\begin{figure}[hbt]
\centering
\includegraphics[width=1\linewidth]{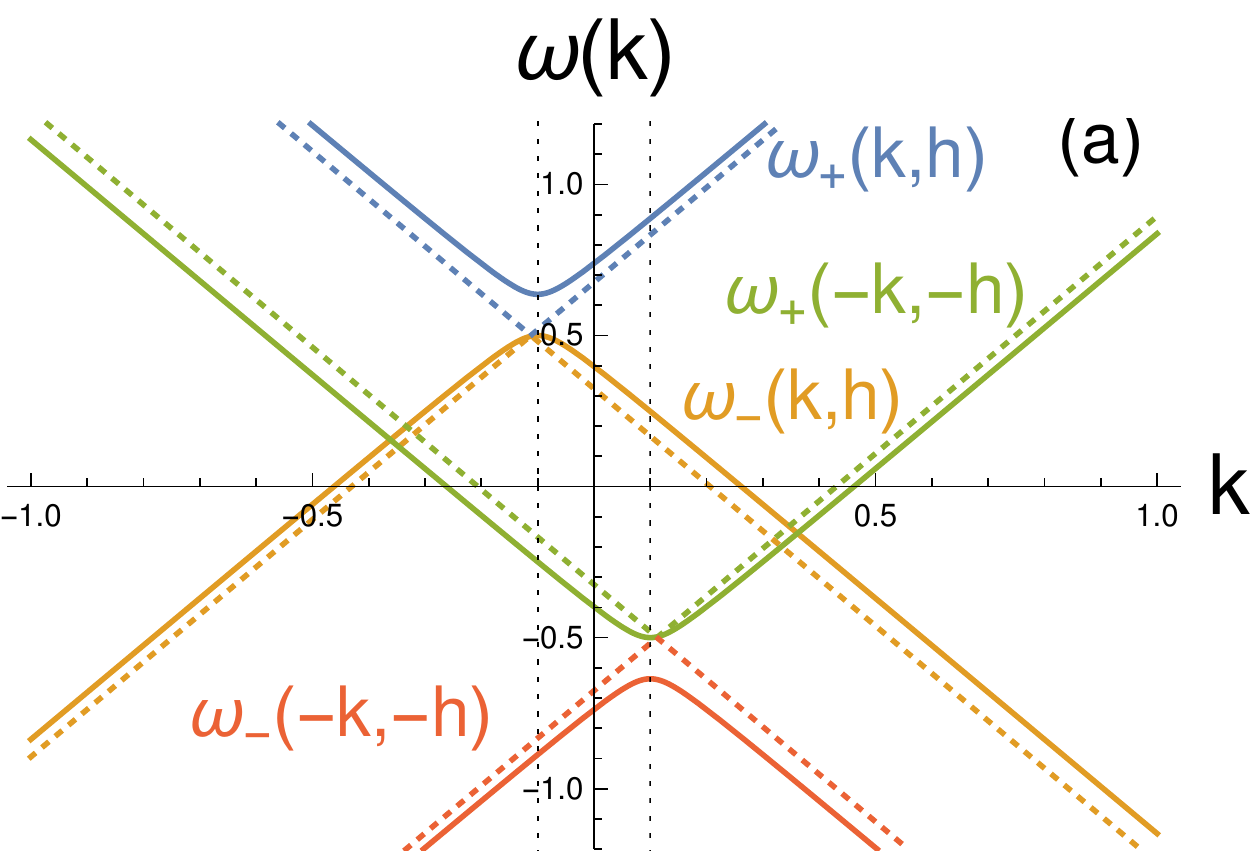}
\includegraphics[width=1\linewidth]{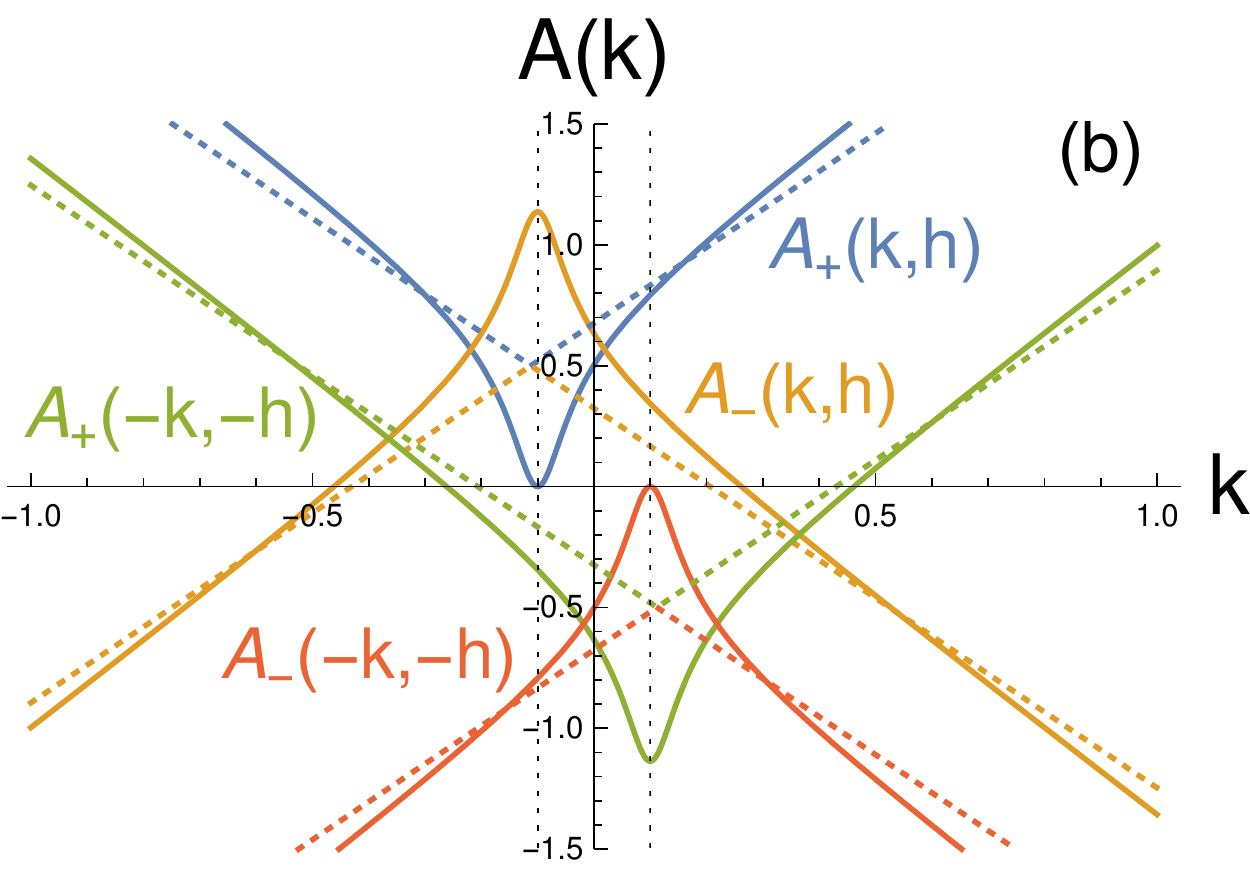}
\caption{(a) The  excitation energies \eqref{5.6} and (b) spectral weights \eqref{5.7} of transverse susceptibilities $\chi^{+-}$ and $\chi^{-+}$ 
for ${\bf h}\parallel {\bf D}$ with $h=0.5$ (in units of $J$).
The dotted vertical line indicates $\pm k_{\rm dm} = \pm D= \pm 0.1$.
Blue and orange indicate modes of $\chi^{+-}$ while green and red indicates those of $\chi^{-+}$.
Solid lines are for $\delta=0.12$ and dashed lines are for $\delta = 0$. }
\label{fig:1}
\end{figure}

It is worth noting that dispersion \eqref{5.6} equally well follows from \eqref{o6} with $k \to \tilde{k}$.

Several features of $\chi^{+-}(k,\omega)$ \eqref{5.3} are worth mentioning. 
The lower branch of excitations, $\omega_{-}(k)$, 
represents the Larmor mode -- its frequency approaches the external Zeeman field $h$ in the limit $\tilde{k} \equiv k + D/J \to 0$, $\omega_{-}(0)=h$, 
while its residue approaches $\chi_0 A_{-}(\tilde{k} = 0) = 2\chi_0 h/(1-\delta) = 2 m$, according to \eqref{4.5}. 
At the same time, the upper branch has higher frequency, $\omega_{+}(\tilde{k} = 0) = (1+\delta)h/(1-\delta)$, but its residue vanishes $A_{+}(\tilde{k} = 0) = 0$. 
Also note that the spin velocity $v$ is renormalized to $\tilde{v} = \sqrt{1-\delta^2} \, v$.

For finite $D \neq 0$, the residue $A_+(k)$ of the upper mode remains finite at $k=0$ (which means $\tilde{k} = k_{\rm dm}= D/J$), as \eqref{5.7} and \eqref{5.8} show.

%Eq.\ \eqref{5.3} was previously derived in Ref.\ \onlinecite{Povarov2021} via a different, fermion path integral based approach, while Ref.\ \onlinecite{Keselman2020} 
%studied ideal spin chain without DM interaction. 

Aside from the momentum shift $k\to k + k_{\rm dm}$, the functional form of Eq.\ \eqref{5.3} coincides with the one derived in Ref.\ \onlinecite{Keselman2020} for the ideal spin chain without DM interaction. It was recently used in Ref.\ \onlinecite{Povarov2022} to explain experimental ESR data in the spin chain with the uniform DM interaction.

Longitudinal spin fluctuations are not affected by the DM in this parallel geometry,
\begin{align}
\chi^{zz}(k,\omega)
&=\frac{\chi_0 \tilde{v} k}{2(1-\delta)} \big(\frac{1}{\omega-\omega_z(k)+i 0^+}-\frac{1}{\omega+\omega_z(k)+i 0^+}\big), \nonumber\\
\omega_{z}(k)&=\sqrt{1-\delta^2} \, v k = \tilde{v} k.
\label{5.5}
\end{align}

Energies of the spin-1 excitations \eqref{5.6} and their respective spectral weights \eqref{5.7} are plotted in Fig.\ \ref{fig:1}. Notice that in agreement with our discussion eigenenergies and their residues of the $\chi^{+-}$ susceptibility are dependent on the combination $k + k_{\rm dm}$ and hence are shifted to the left along the $k$-axis, while those of the $\chi^{-+}$ susceptibility depend on $k - k_{\rm dm}$ and are shifted in the opposite direction, to the right.

\subsection{${\bf h}\perp{\bf D}$}
\label{sec:DperpH}

For ${\bf h}\perp{\bf D}$, we set $\tilde{\bf D}=\tilde{D}\hat{\bf x}$ so that $\tilde{D}^z = 0, \tilde{D}^\pm = \tilde{D}$ in \eqref{4.11}. Accordingly, the spin current develops finite expectation value $j^\pm = \chi_0 \tilde{D}/(1+\delta)$ but $j^z =0$. The problem lacks any continuous spin symmetry and transverse and longitudinal fluctuations are now coupled.

\begin{figure}[hbt]
\centering
\includegraphics[width=1\linewidth]{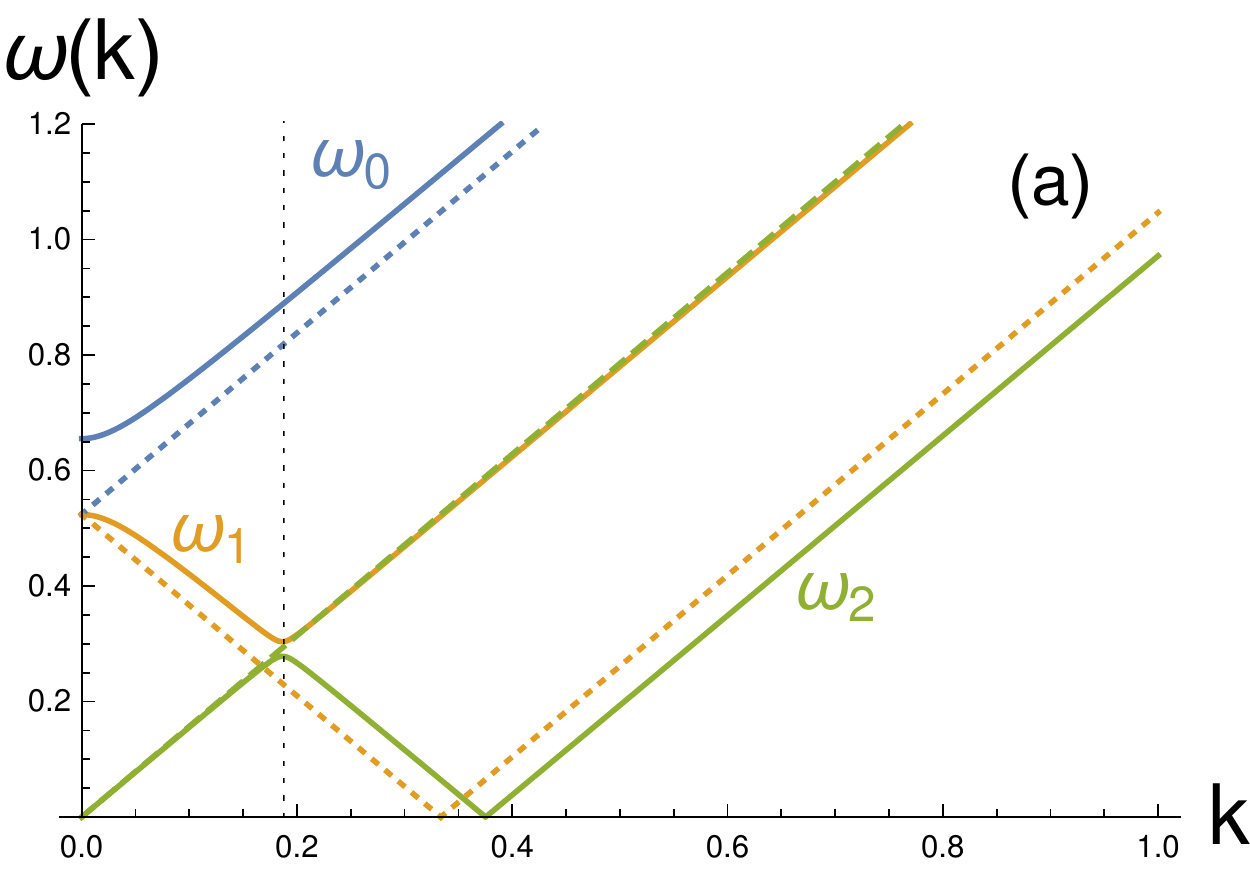}
\includegraphics[width=1\linewidth]{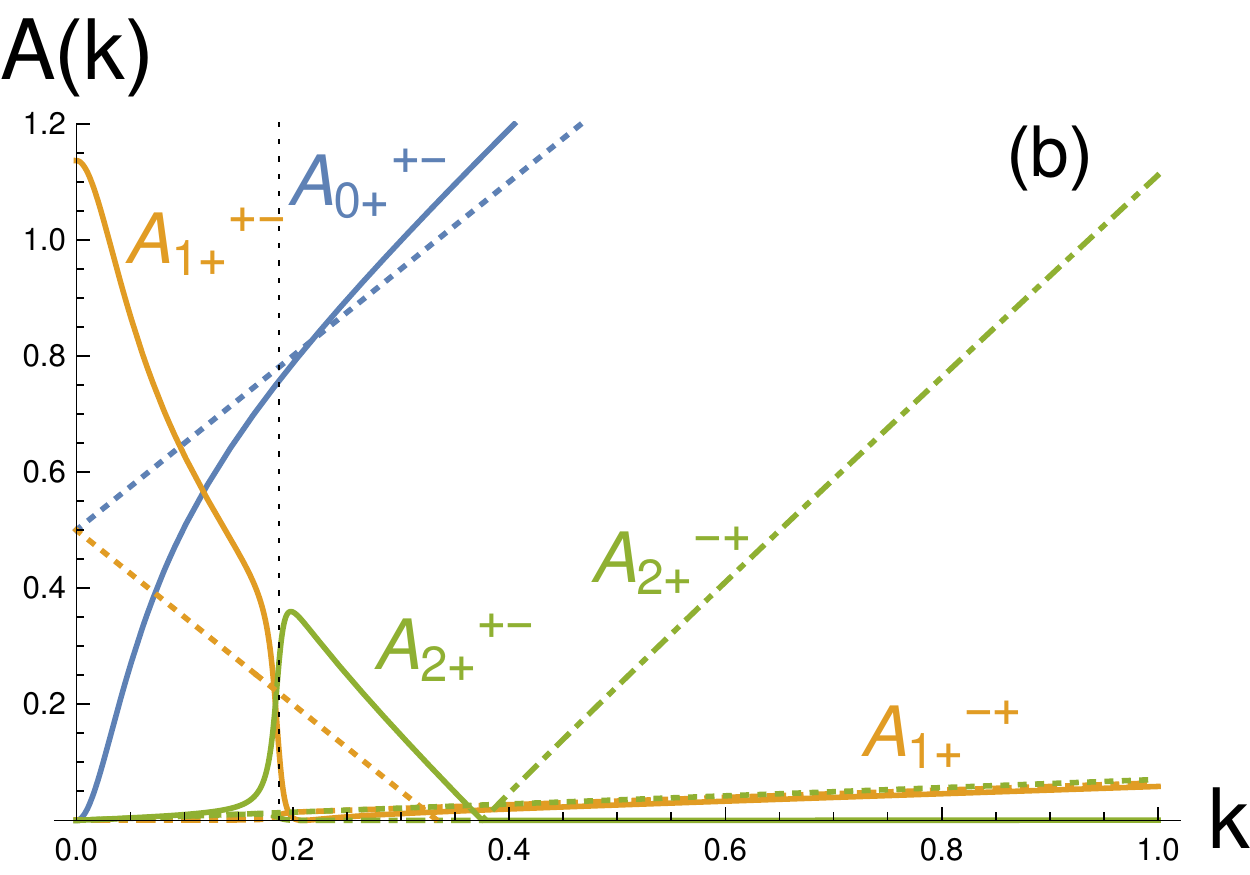}
\caption{
(a) The  excitation energies \eqref{A.2} and (b) spectral weights of transverse susceptibilities $\chi^{+-}$ and $\chi^{-+}$ [\eqref{A.6}] for ${\bf h}\perp {\bf D}$ with $h=0.5$ 
and $D=0.1$ (in units of $J$).
Blue, orange, and green indicate modes $\omega_{0}$, $\omega_{1}$, and $\omega_{2}$, respectively.
Solid and dotted-dash lines are for $\delta=0.12$ and dotted lines are for $\delta= 0$.
The dotted vertical line indicates $vk=B/2$, where $B$ is the total magnetic field \eqref{jan1}.
We find that  $A_{0+}^{-+}$ is about $10^3$ times smaller than $A_{0+}^{+-}$  and do not plot it in the figure.}
\label{fig:2}
\end{figure}

\begin{figure}[hbt]
\centering
\includegraphics[width=1\linewidth]{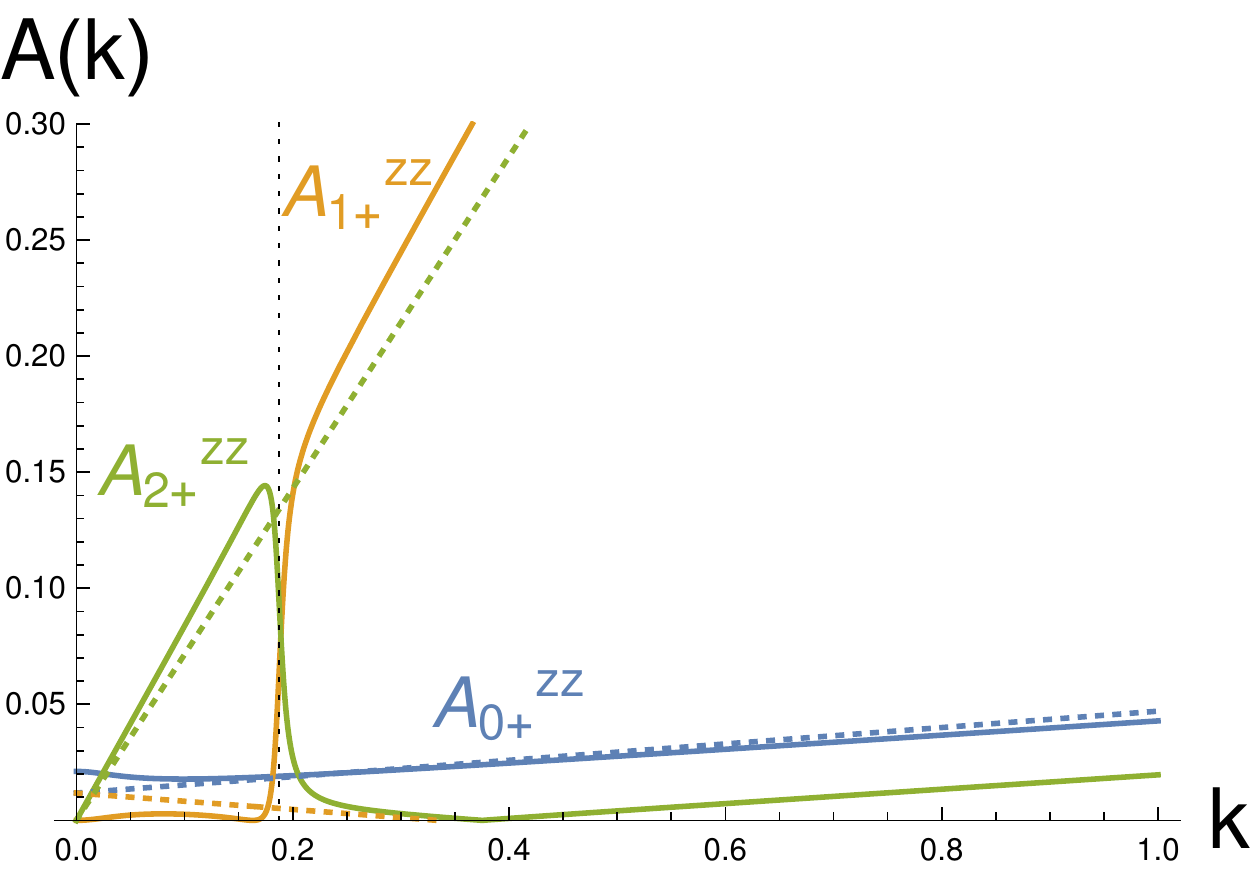}
\caption{
The spectral weights of longitudinal  susceptibility $\chi^{zz}$ \eqref{A.7} for ${\bf h}\perp {\bf D}$ with $h=0.5$ and $D=0.1$ (in units of $J$).
Blue, orange, and green indicate modes $\omega_{0}$, $\omega_{1}$, and $\omega_{2}$, respectively.
Solid lines are for $\delta=0.12$ and dotted lines are for $\delta= 0$.
The  dotted vertical line indicates $vk=B/2$.}
\label{fig:3}
\end{figure}

Solving the characteristic equation
\begin{align}
\det(\omega-{\cal A}(k))=0,
\label{5.9}
\end{align}
we find excitation energies $\omega_i(k)$, where $i=0,1,2$. 
It is actually possible to solve the matrix equation \eqref{5.2} analytically and details are provided in Appendix \ref{app:a}.
Extensive algebraic manipulations of \eqref{5.2} lead to 
\begin{align}
\chi^{ab}(k,\omega) 
= \sum_{i=0}^2 \sum_{\eta = \pm} \frac{A^{ab}_{i \eta}(k)}{\omega - \eta \, \omega_i(k) + i0^+},
\label{ap101}
\end{align}
where $a,b=+,-,z.$
These results are illustrated in Figs.\ \ref{fig:2} and \ref{fig:3} which plot excitation energies $\omega_i$ in \eqref{A.2} of the spin-1 excitations and their respective spectral weights  $A_{i+}^{+-}$ \eqref{A.6} and $A_{i+}^{zz}$ \eqref{A.7} as a function of momentum $k$.

With the goal of understanding the ESR experiments, here we present relevant spin susceptibilities at $k=0$. We find 
$
A_{1\pm}^{zz}(0)
=A_{2 \pm}^{zz}(0)
=A_{0\pm}^{+-}(0)
=A_{2\pm}^{+-}(0)
=0,
$
as can also be seen from Figs.\ \ref{fig:2} and \ref{fig:3}, and 
\begin{align}
\chi^{+-}(0,\omega)
=&\chi_0[\frac{A_{1+}^{+-}(0)}{\omega-\omega_1(0) + i0^+}
+\frac{A_{1-}^{+-}(0)}{\omega+\omega_1(0) + i0^+}],\label{5.10}\\
\chi^{-+}(0,\omega)
=&\chi_0[\frac{-A_{1-}^{+-}(0)}{\omega-\omega_1(0)+i0^+}+\frac{-A_{1+}^{+-}(0)}{\omega+\omega_1(0) + i0^+}],\label{5.11}\\
\chi^{zz}(0,\omega)
=&\chi_0[\frac{A_{0+}^{zz}(0)}{\omega-\omega_0(0) + i0^+}-\frac{A_{0+}^{zz}(0)}{\omega+\omega_0(0) + i0^+}]. \label{5.12}
\end{align}
Close similarity between transverse susceptibilities $\chi^{+-}$ and $\chi^{-+}$ is the consequence of the Onsager's relation.
Spin excitation energies at $k=0$ are given by 
\begin{align}
\omega_0(0)
&=\sqrt{\big(\frac{1+\delta}{1-\delta}\big)^2 h^2+\frac{1-\delta}{1+\delta}\tilde{D}^2},\label{5.14}\\
\omega_{1}(0)
&=\sqrt{h^2+\frac{1-\delta}{1+\delta}\tilde{D}^2},\label{5.13}\\
\omega_{2}(0) &= 0 , \label{o28}
\end{align}
and the residues are  
\begin{align}
A_{1\pm}^{+-}(0)
&=\frac{h}{1-\delta}\pm\frac{\frac{h^2}{1-\delta}+\frac{1}{2}\frac{\tilde{D}^2}{1+\delta}}{\omega_{1}(0)},\quad
A_0^{zz}(0)
=\frac{1}{2}\frac{\frac{\tilde{D}^2}{1+\delta}}{\omega_0(0)}.
\label{o8}
\end{align}

We observe that at $k=0$ there is a single pole in $\chi^{+-}$ -- the system responds at the frequency 
\be
\omega_1(0)=\sqrt{h^2 + (1-\delta^2)(v D/J)^2} > h ,
\label{o26}
\ee
where we used \eqref{5.8} for $\tilde{D}$.

The absence of $\omega_0(0)$ \eqref{5.14} in $\chi^{+-}(0,\omega)$ follows from the geometry of the problem, ${\bf h}\perp{\bf D}$, and is specific to $k=0$ limit. A short manipulation of \eqref{4.7} and \eqref{4.8} 
with $k=0$, ${\bf h} = (0,0,h)$ and $\tilde{\bf D} = (\tilde{D},0,0)$ shows that six linear equations \eqref{4.9} factorize into two groups of three equations each.  The first of these `triplets' describes coupled motion of 
$(\delta \hat m^+,\delta \hat m^-, \delta \hat j^z)$,
\be
\begin{pmatrix}
\omega-h &0& \tilde{D}\\
0&\omega+h & -\tilde{D}\\
\frac{(1-\delta)\tilde{D}}{2(1+\delta)} & -\frac{(1-\delta)\tilde{D}}{2(1+\delta)} & 0
\end{pmatrix}
\begin{pmatrix}
\delta \hat m^+\\
\delta \hat m^-\\
\delta \hat j^z
\end{pmatrix} = 0.
\label{o31}
\ee
It is solved by $\omega=0$, which is \eqref{o28}, and $\omega = \pm \omega_1(0)$ \eqref{5.13}. This explains the absence of the resonant response of $\chi^{+-} \sim \langle \delta \hat m^+ \delta \hat m^- \rangle$ at the frequency $\omega_0(0)$, \eqref{5.14}.

The second group is made of $(\delta \hat j^+,\delta \hat j^-, \delta \hat m^z)$ and is described by 
\be
\begin{pmatrix}
\omega-\frac{(1+\delta)h}{1-\delta}  &0& \frac{(1-\delta)\tilde{D}}{1+\delta}\\
0&\omega + \frac{(1+\delta)h}{1-\delta}&-\frac{(1-\delta)\tilde{D}}{1+\delta}\\
\frac{1}{2}\tilde{D} & -\frac{1}{2}\tilde{D} & 0
\end{pmatrix}
\begin{pmatrix}
\delta \hat j^+\\
\delta \hat j^-\\
\delta \hat m^z
\end{pmatrix} = 0.
\label{o32}
\ee
It is solved by $\omega=0$ and $\omega = \pm \omega_0(0)$, \eqref{5.14}. The mixing of $\delta \hat j^\pm$ spin currents with longitudinal magnetization fluctuations $\delta \hat m^z$ explains why the longitudinal susceptibility $\chi^{zz} \sim \langle \delta \hat m^z \delta \hat m^z \rangle$ responds at $\omega = \omega_0(0)$ but not at $\omega_1(0)$.

Equations \eqref{4.7} and \eqref{4.8} show that at finite $k\neq 0$ these two groups of spin fluctuations hybridize, leading to complicated evolution of the dispersions and the spectral weights at finite $k$, shown in Figs.\ \ref{fig:2} and \ref{fig:3}. It is worth adding that spin susceptibilities in Figs.\ \ref{fig:2} and \ref{fig:3} also possess an interesting avoided level crossing between $\omega_1$ and $\omega_2$ branches at the momentum 
$k_0=B/(2v)$, where 
\be
B=\sqrt{\big(\frac{h}{1-\delta}\big)^2+\big(\frac{\tilde{D}}{1+\delta}\big)^2} = \sqrt{\big(\frac{h}{1-\delta}\big)^2+\big(\frac{v D}{J}\big)^2},
\label{jan1}
\ee
represents the total magnetic field, the sum of the external and internal molecular fields, experienced by spinons.
The splitting between two branches is found from the general expressions in Appendix \ref{app:a} to be
$\pi D h \delta/\Big(\sqrt{2} \sqrt{h^2 + (v D/J)^2}\Big)$ and is therefore due to the combined effect of finite $D$, $h$ and the interaction $g_\mathrm{bs}$. 
Its experimental observation requires high-precision measurements at finite momenta.

The fact that  both $\omega_1(0)$ and $A_{1+}^{+-}(0)$ remain finite even in the $h\to0$ limit implies that finite energy absorption rate is present even without the applied external field, in agreement with earlier experimental observations and the non-interacting spinon theory \cite{Povarov2011}.

Broken spin-rotational symmetry leads to the finite absorption in the longitudinal sector, $\chi^{zz}(0,\omega)$,  as well. It takes place at the higher frequency $\omega_0(0)$, that is distinct from the spin-current frequency $\omega_+(k=0)$ of the previous Section \ref{sec:h||d}. The residue of this signal $A_0^{zz}(0)$ is finite, but its observation requires Voigt geometry when the microwave field is polarized along the direction of the external field $h$.

Notice that in the $h\to 0$ limit $\omega_0(0) = \omega_1(0)$ and hence the residues coincide too, $A_0^{zz}(0) = A_{1+}^{+-}(0)$, see \eqref{o8}. This is the case of zero-field absorption when $\chi^{zz}(0,\omega)|_{h=0}$ and $\chi^{+-}(0,\omega)|_{h=0}$ describe transverse, with respect to the `built-in' DM field $\tilde{\bf {D}}$, response that is coupled linearly to the magnetization current $\hat{\bf {J}}$ and is oriented along the $\hat{\bf x}$ axis.

\subsection{Arbitrary angle $\theta$ between ${\bf h}$ and ${\bf D}$}
\label{sec:arbDH}

We choose ${\bf D}$ and ${\bf h}$ to be in the xz-plane, set $\tilde{\bf D}=\tilde{D}(\sin\theta\hat{\bf x}+\cos\theta\hat{\bf z})$ and focus on analyzing the uniform dynamic susceptibility $\chi^{ab}(k=0,\omega)$ below.
At $k=0$, the eigenvalues of \eqref{4.9} are given by the simple expression
\bea
\Omega^2_{\mu = \pm}(\theta)
&=& \frac{1-\delta }{1+\delta }\tilde{D}^2 \sin^2(\theta) + \Big( \frac{h}{1-\delta} \nonumber\\
&& + \mu \sqrt{\big(\frac{h \delta}{1-\delta}\big)^2 + \frac{1-\delta }{1+\delta }\tilde{D}^2 \cos^2(\theta)} \,\, \Big)^2 .
\label{5.15}
\eea
This expression can be understood as a result of the hybridization between the positive and negative frequency branches of $\delta \hat m^\pm$ and $\delta \hat j^\pm$ fluctuations 
with $\delta \hat j^z$ and $\delta \hat m^z$ modes, correspondingly. This hybridization is mediated by $\tilde{D}^\pm=\tilde{D} \sin(\theta)$ terms in \eqref{4.11}. 
%and $j^\pm \sim \tilde{D} \sin(\theta)$ ones in \eqref{o36}.

Eq.\ \eqref{5.15} is seen to interpolate between $\omega_\pm(k=0)$ in \eqref{5.6} for $\theta=0$, for the case of ${\bf h}\parallel {\bf D}$, to $\omega_{0,1}(0)$ in \eqref{5.14} and \eqref{5.13} for the ${\bf h}\perp{\bf D}$ case, when $\theta=\pi/2$. It is easy to see that for $\theta > 0$ these energies are finite, $\Omega_\pm \neq 0$, as long as $D \neq 0$.

The $k=0$ but $\theta$-dependent $\chi^{ab}(k=0,\omega;\theta) \equiv \chi^{ab}(\omega;\theta)$ dynamic susceptibility is found to be 
\be
\chi^{ab}(\omega;\theta) =\chi_0\sum_{\mu=\pm}\sum_{\eta=\pm}\frac{{\tilde A}_{\mu \eta}^{ab}(\theta)}{\omega-\eta \Omega_\mu(\theta)+i0^+},
\label{o33}
\ee
where $a,b=+,-,z$.
Similar to dispersions $\Omega_\pm(\theta)$, spectral weights ${\tilde A}_{\mu \eta}^{ab}(\theta)$ interpolate from \eqref{5.7} at $\theta=0$ to \eqref{o8} at $\theta=\pi/2$. Their explicit forms are listed in \eqref{B.3} and \eqref{B.4} and plotted, together with \eqref{5.15}, in Figs.\ \ref{fig:4} and \ref{fig:5} {\em vs.} angle $\theta$. 
More details are in Appendix \ref{app:b}.

\begin{figure}[hbt]
\centering
\includegraphics[width=1\linewidth]{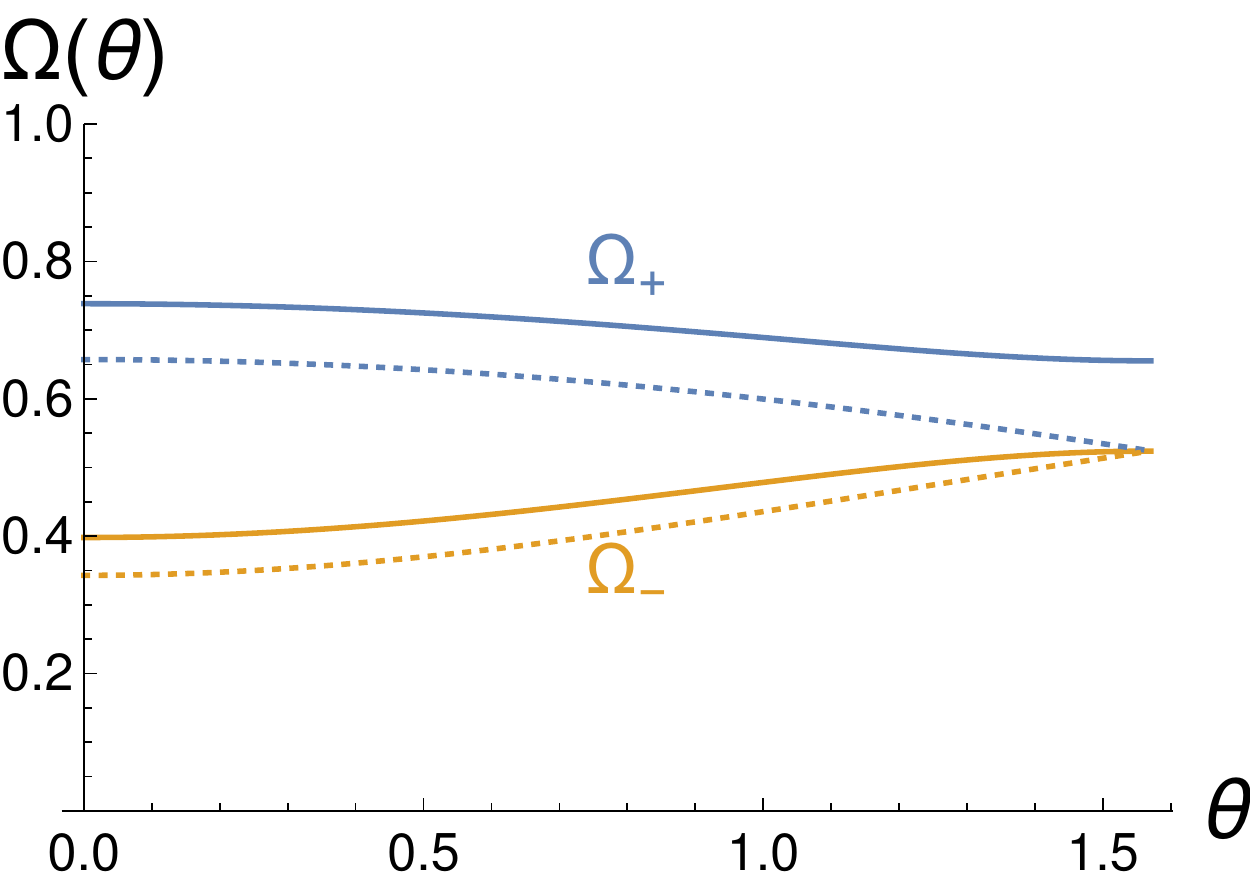}
\caption{
The excitation energies  of   susceptibilities  $\chi^{+-}$, $\chi^{-+}$, and  $\chi^{zz}$ as a function of $\theta$   for the case ${\bf h}$ in arbitrary directions with ${\bf D}$ with  $h=0.5$ 
and $D=0.1$ (in units of $J$).
Blue and  orange indicate modes $\Omega_{+}$ and $\Omega_{-}$, respectively.
Solid lines are for $\delta=0.12$ and dotted lines are for $\delta= 0$.}
\label{fig:4}
\end{figure}

Fig.\ \ref{fig:4} shows that the splitting between $\Omega_{+}$ and $\Omega_{-}$ is finite for all $\theta$, in a contrast to the non-interacting, $\delta=0$, situation for which the dispersions are shown 
by the dashed lines.
In that case the splitting $\propto \tilde{D} \cos(\theta)$ and vanishes in the orthogonal configuration $\theta=\pi/2$.  \cite{Povarov2011}

This quantitative difference between $\delta \neq 0$ and $\delta=0$ situations is, however, partially compensated by the nontrivial evolution of spectral weights ${\tilde A}^{+-}_{\pm +}$ with the angle, 
as illustrated in Fig.\ \ref{fig:5}(a). 
There, one observes that the spectral weight ${\tilde A}^{+-}_{+ +}(\theta)$ of the upper mode $\Omega_{+}(\theta)$ actually vanishes at $\theta = \pi/2$. That is, similar to the 
non-interacting case, for $\theta=\pi/2$ there is only one resonance frequency $\Omega_{-}(\pi/2)$ in the transverse dynamic susceptibility $\chi^{+-}(\omega;\pi/2)$.

Fig.\ \ref{fig:5}(b) shows that at the same time the longitudinal susceptibility $\chi^{zz}(\omega;\pi/2)$ demonstrates complimentary behavior. Here, the only resonant frequency present at $\theta=\pi/2$ 
is $\Omega_{+}(\pi/2)$ because the spectral weight ${\tilde A}^{zz}_{-+}(\theta)$ of the $\Omega_{-}(\pi/2)$ pole vanishes at $\theta=\pi/2$. 

Both of these features are special to ${\bf h} \perp {\bf D}$ and $k=0$ limits and are explained in the preceding Section \ref{sec:DperpH}, see equations \eqref{o31}, \eqref{o32} and discussion there.
 
\begin{figure}[hbt]
\centering 
\hspace*{.01cm}
\includegraphics[width=.97\linewidth]{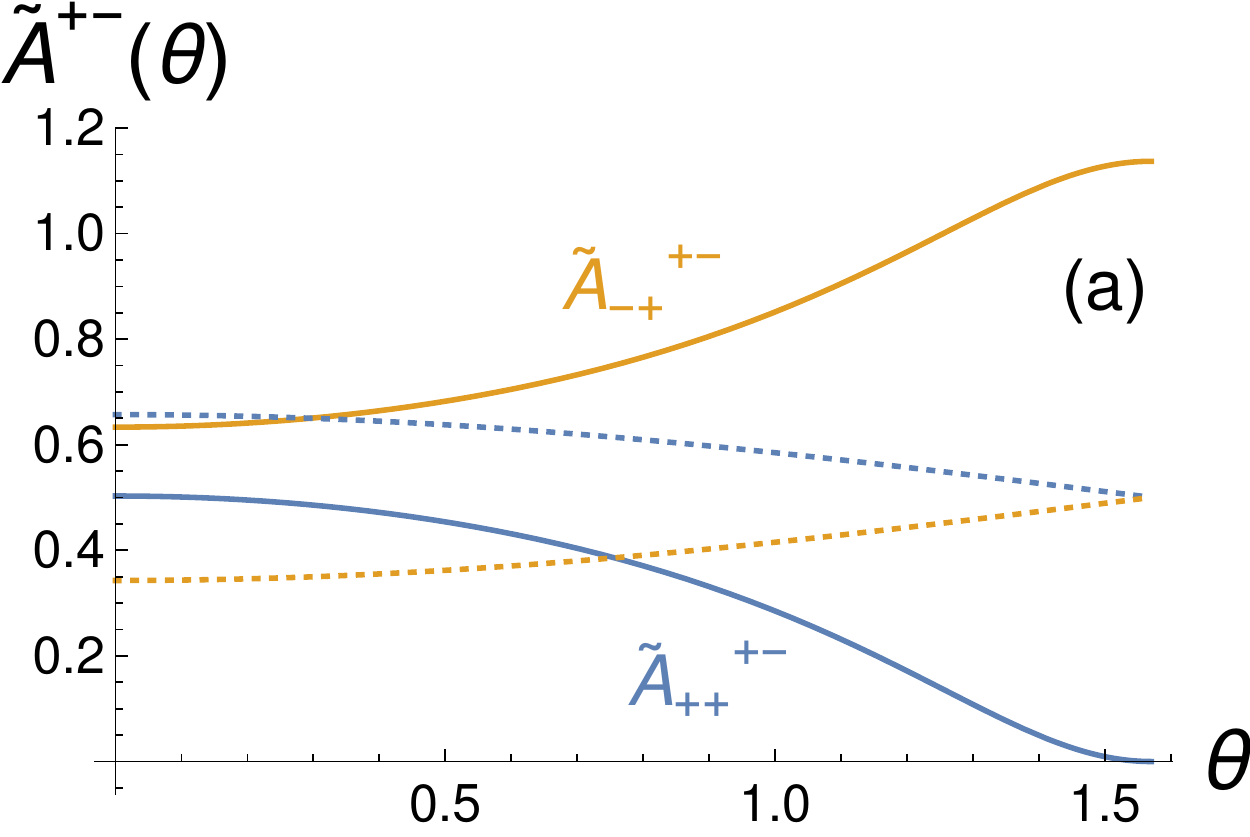}
\includegraphics[width=1\linewidth]{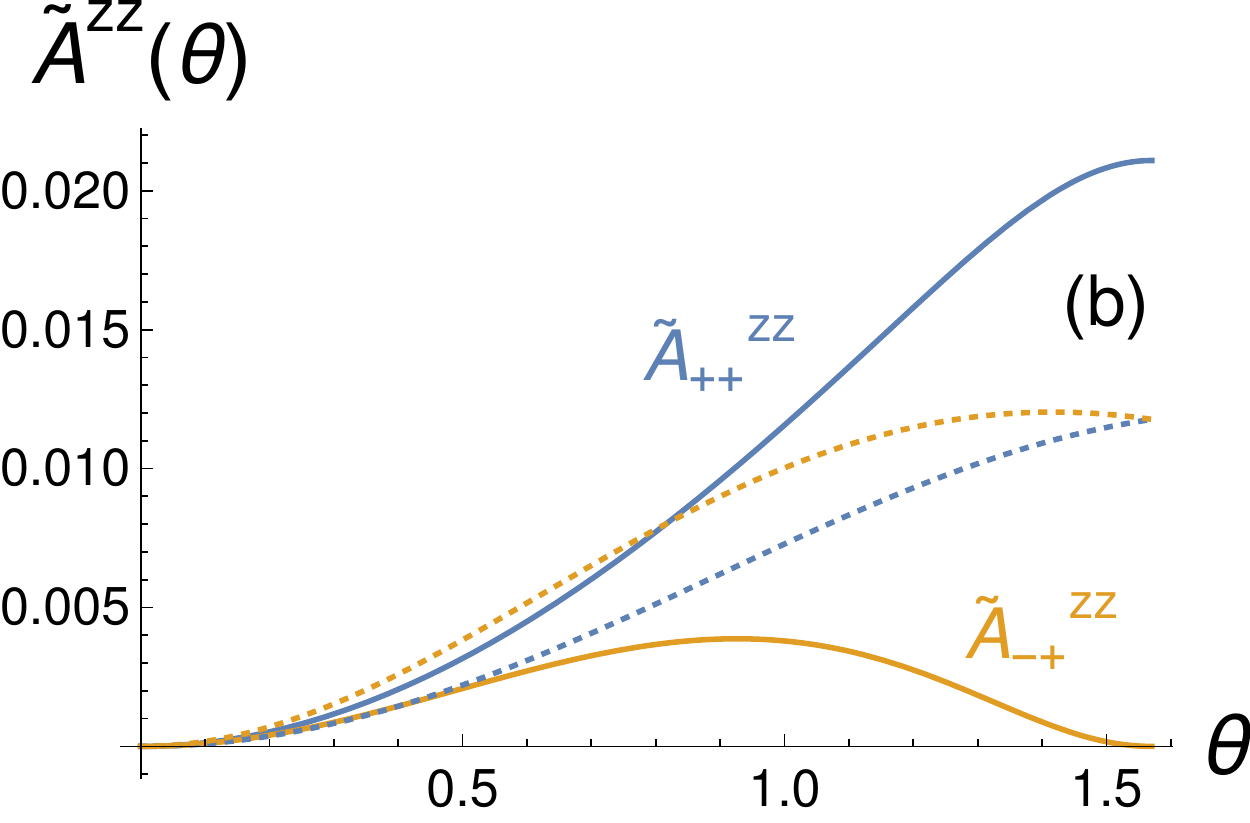}
\caption{
The spectral weights of   susceptibilities (a)  $\chi^{+-}$  and (b) $\chi^{zz}$ as a function of $\theta$   for the case ${\bf h}$ in arbitrary directions with ${\bf D}$ with  $h=0.5$ 
and $D=0.1$ (in units of $J$).
Blue and  orange indicate modes $\Omega_{+}$ and $\Omega_{-}$, respectively.
Solid lines are for $\delta=0.12$ and dotted lines are for $\delta= 0$.}
\label{fig:5}
\end{figure}

\section{Interaction effect on the electron spin resonance}
\label{sec:esr}

Electron spin resonance is a uniquely sensitive probe of the spin dynamics at $k=0$ and is particularly well suited for probing physics described in this paper, as was convincingly demonstrated previously 
\cite{Povarov2011,Smirnov2015,Povarov2022}. Within the linear response theory, the rate of the energy absorption per unit length, which is measured by ESR, is given by the intensity
\begin{align}
I(\omega)
=-\frac{1}{2}H_\mathrm{rad}^2\omega \Im\chi^{nn}(k=0,\omega),
\label{os34}
\end{align}
where $H_\mathrm{ rad}$ is the amplitude of the radiation (microwave) field that the sample is radiated with. In the continuum limit this is described by the monochromatic perturbation 
$V(t) = - \int dx H_\mathrm{rad} e^{-i\omega t} {\bf n} \cdot \hat{\bf M}(x)$ and ${\bf H}_\mathrm{ rad}$ is linearly polarized along the direction ${\bf n}$.
In the frequently employed Faraday geometry ${\bf n}$ is chosen to be in the plane normal to the static field ${\bf h}$. 
For example, for ${\bf n} = {\bf x}$ the rate of absorption is controlled by the spin-flip processes and is determined by 
$[\Im\chi^{+-}(k=0,\omega) + \Im\chi^{-+}(k=0,\omega) +  \Im\chi^{++}(k=0,\omega) +  \Im\chi^{--}(k=0,\omega)]/4$. 
(Typically, contributions from $\chi^{\pm \pm}(k=0,\omega)$ are very small, their spectral weight $\propto D^2$.)
As noted previously, to probe longitudinal susceptibility $\Im\chi^{zz}(k=0,\omega)$, one needs to use Voigt geometry when ${\bf n}$ is directed along the external field ${\bf h}$.

\begin{figure}[hbt]
\centering
\includegraphics[width=1\linewidth]{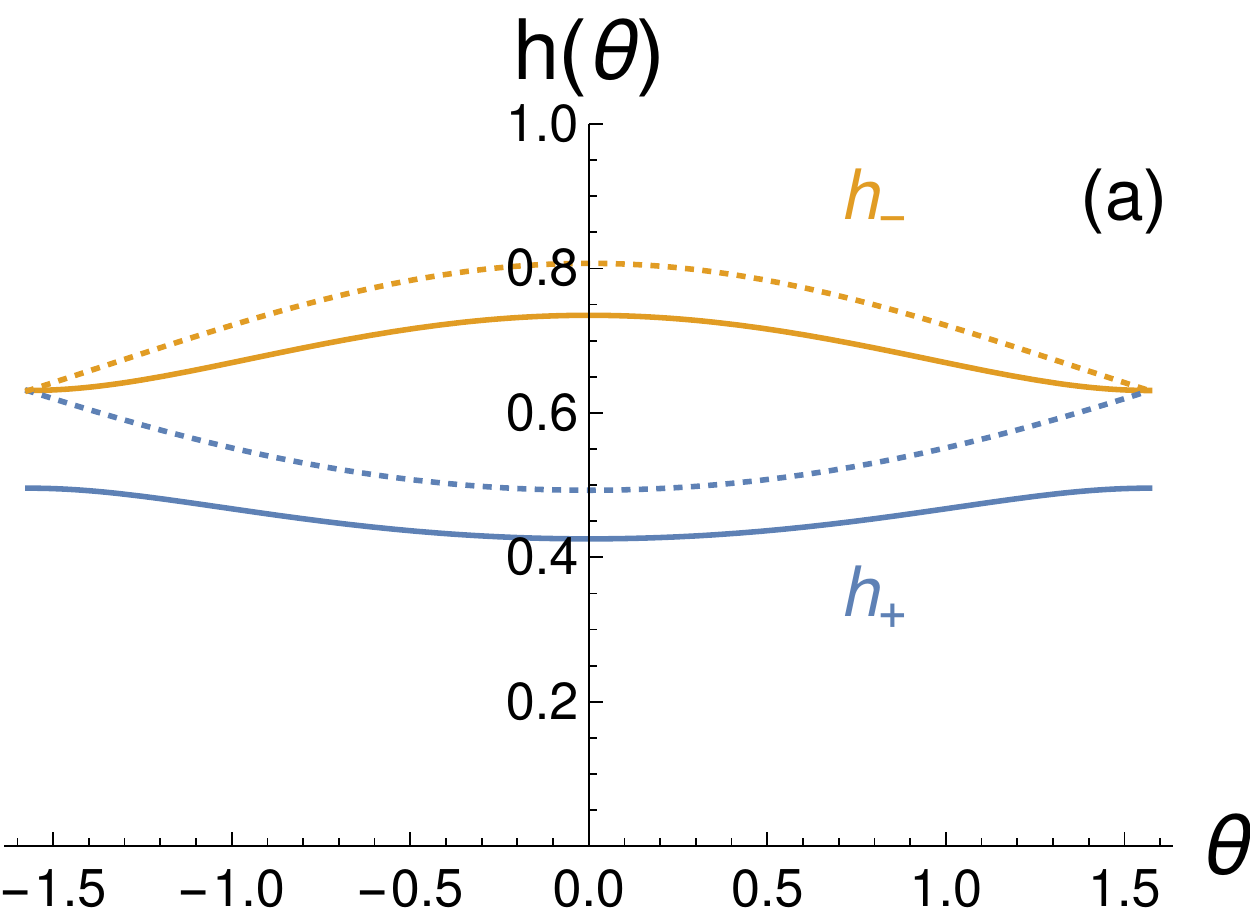}
\includegraphics[width=1\linewidth]{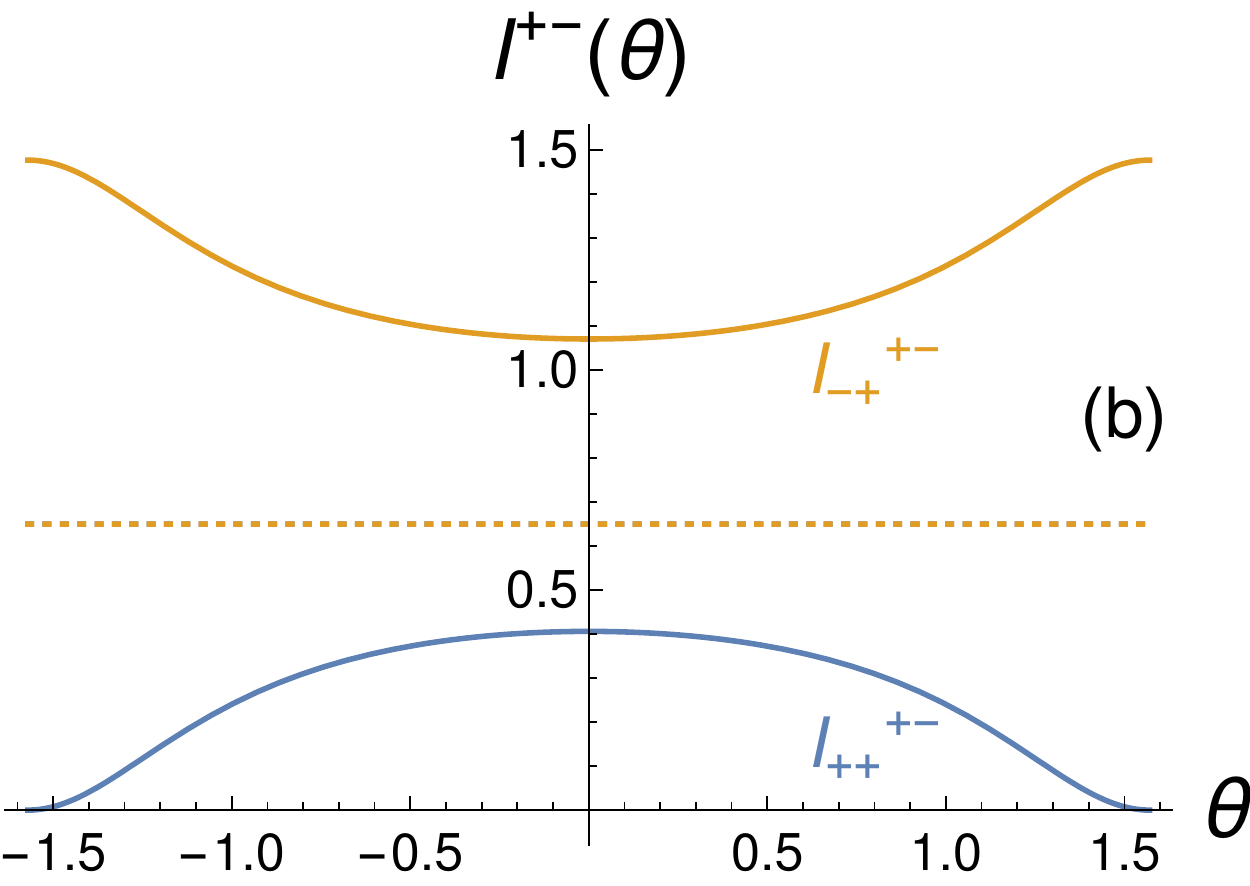}
\caption{
The resonant fields $h_\pm(\theta)$ and intensities $I^{+-}_{\pm+}$ of   susceptibilities  $\chi^{+-}$ as a function of $\theta$   for the case ${\bf h}$ in arbitrary directions with ${\bf D}$ with  $h=0.5, D=0.1$, and $\omega=0.65$ (in units of $J$).
Blue and  orange indicate modes $\Omega_{+}$ and $\Omega_{-}$, respectively.
Solid lines are for $\delta=0.12$ and dotted lines are for $\delta= 0$.
Note that the non-interacting intensities for the resonant fields  are the same and constant.}
\label{fig:6}
\end{figure}

In addition, actual ESR measurements are done at the fixed frequency $\omega$, specific to the resonant cavity in which the sample is held, as a function of varying magnetic field $h$. Given \eqref{5.15}, 
the resonant fields $h_\pm$ corresponding to $\Omega_{\pm}(\theta)$ are  
\bea
h_\pm(\theta) 
&=& \Big(\sqrt{\omega^2 - \tilde{v}^2 d^2 \sin^2\theta} + \\
&&\mp \sqrt{\delta^2 (\omega^2 - \tilde{v}^2 d^2) + \tilde{v}^2 d^2 \cos^2\theta} \Big)/(1+\delta), \nonumber
\label{os35}
\eea
where we used \eqref{5.8} for $\tilde{D}$ and \eqref{5.5} for $\tilde{v}$, and abbreviated $d = D/J$. 
Observe that the excitation frequency is bounded from below by 
$\omega = \tilde{v} d \approx \sqrt{1-\delta^2} \pi D/2$, which is just \eqref{5.15} in the case of the vanishing magnetic field $h=0$. Figure \ref{fig:6}(a) shows $h_\pm(\theta)$ 
for the specific choice of parameters $D=0.1, \delta=0.12, \omega = 0.65$, in units of exchange interaction $J$.

Using \eqref{o33}, the intensity as a function of $\omega$ is 
\be
I(\omega)
=\frac{\pi}{2}H_\mathrm{rad}^2 \chi_0 \omega \sum_{\mu=\pm}\sum_{\eta=\pm} {\tilde A}_{\mu \eta}^{ab}(\theta) \delta(\omega-\eta \Omega_\mu(\theta)).
\label{os37}
\ee
To write it as a function of the external field $h$, we need to `solve' the delta function by using $\Omega_\mu(h) = \Omega_\mu(h_\mu) + (h-h_\mu) \Omega_\mu' $, where 
$\Omega_\mu' = (d\Omega_\mu/d h)|_{h=h_\mu}$. Note that by construction $\Omega_\mu(h_\mu) = \omega$. 
Then $\delta(\omega - \Omega_{\mu=\pm}) = \delta(h - h_{\mu})/|\Omega_\mu'|$ and one obtains
\be
I(h) =  \sum_{\mu=\pm} I_{\mu +}^{ab}(h,\theta) \delta(h - h_\mu(\theta))
\label{os38}
\ee
where partially intensities $I_{\mu +}^{ab}$ describe contributions originating from modes $\Omega_\mu$ ($\mu = \pm$) of the dynamic spin susceptibility $\chi^{ab}$, with $a,b=(+,-,z)$.

Fig.\ \ref{fig:6}(b) shows the so obtained intensities at the resonant field $h_{+}$, $(\mu,\eta)=(+,+)$, and $h_{-}$, $(\mu,\eta)=(-,+)$ of the transverse susceptibility $\chi^{+-}$, $(a=+, b= -)$. 
Being interested in relative intensities, we set $\pi H_\mathrm{rad}^2 \chi_0 \omega/2 = 1$ in the plot.
In agreement with the discussion in the previous section we observe the upper mode intensity $\propto \tilde{A}_{++}^{+-}(\theta)$ to vanish in the orthogonal configuration $\theta=\pi/2$. Therefore, for this specific angle there is only one resonance, at the field $h_{-}$. 
For all other values of the angle between ${\bf h}$ and ${\bf D}$, there are two resonances, at fields $h_{+}$ and $h_{-}$. 
Fig.\ \ref{fig:6}(b) shows that intensity of the $h_{-}$ resonance is generally greater than that of the $h_{+}$ one. 
We believe this simple feature of our theory explains experimental data on the angular dependence of modes $M_{-}$ and $M_{+}$, presented in Fig.\ 8 of Ref.\ \onlinecite{Smirnov2015}. 
It is seen there that mode $M_{+}$, that is the signal at the resonant field $h_{+}$ corresponding to the upper mode $\Omega_{+}(\theta)$, is observed only within a finite angular interval of (approximately) $\theta \leq \pi/3$. The explanation is that the intensity of this mode falls below experimentally detectable value for bigger $\theta$.

Another notable feature of \eqref{os38} is that generally $I_{\mu +}^{+-}$ is the biggest. This is easy to understand by recalling that in the absence of the DM interaction the only susceptibility that contributes to the ESR is $\chi^{+-}$. However, for finite $D$ and relatively small angles between ${\bf h}$ and ${\bf D}$, 
$\theta \leq \pi/4$, there also is a noticeable contribution from $\chi^{-+}$ susceptibility, especially for small magnetic field $h \approx D$. 
This contribution is most prominent in the parallel configuration, $\theta=0$, and has been observed experimentally in Ref.\ \onlinecite{Povarov2022}. 
Relative smallness of this contribution is a consequence of the small $D/J$ ratio -- the $k=0$ signal from $\chi^{-+}$ is present only because the spin-rotational symmetry of the chain is broken by the DM interaction. 

More extended discussion of this and other features of the theory relevant to modern ESR experiments are presented in Appendix \ref{app:b}.

We conclude this section with a brief comparison of the non-interacting spinon description of the DM-induced ESR doublet \cite{Povarov2011} with the more complete interacting spinon theory presented here and, for the parallel configuration $\theta=0$, in Ref.\ \onlinecite{Povarov2022}. 
Within the former description, the splitting between $\Omega_\pm$ modes vanishes for $\theta=\pi/2$. 
As a result, the double resonance reduces to the single one (two contributions at the same frequency/resonant field) \cite{Povarov2011}. 
For the interacting spinons the splitting is always finite, see Fig.\ \ref{fig:6}(a). 
But the relative intensity of the two contribution varies greatly with the relative angle between the field and the DM vector, and vanishes in the orthogonal configuration as Fig.\ \ref{fig:6}(b) shows. Therefore, $\Omega_{+}$ remains distinct from $\Omega_{-}$, but its spectral weight disappears at $\theta=\pi/2$. Therefore, in both considerations, only one resonance is present at $\theta=\pi/2$.

\section{Numerical simulations}
\label{sec:dmrg}

We now compare our analytical predictions with numerical simulations using matrix-product-state techniques.
Our numerical calculations are carried out using the ITensor library~\cite{ITensor}.
To obtain the spectral function~\eqref{o2} we first obtain the ground state of the system, $|\Psi_{\rm gs}\rangle$ using density matrix renormalization group (DMRG)~\cite{White1992}. We then perform time evolution of the quenched state $\hat{S}_0^-|\Psi_{\rm gs}\rangle$ (where $n=0$ corresponds to a site in the middle of the chain) up to times $t_{\rm max}=40 J^{-1}$. To this end we use time evolving block decimation (TEBD)~\cite{Vidal2004} employing a 4th order Suzuki-Trotter decomposition with a time step of $dt=5 \cdot 10^{-3}$.  
Our analysis is done on finite systems of length $N=200$ sites with open boundary conditions. 
%To reduce numerical errors we perform averaging of the real-space spin-spin correlations using simulations with $\vec{D}\to - \vec{D}$ using correlations with $n\to -n$.
Employing the symmetry of the Hamiltonian upon inversion of the DM interaction vector ${\bf D} \to -{\bf D}$ followed by spatial inversion, we perform a symmetrization of the real-space spin-spin correlations using simulations carried out for both DM orientations. %, which reduces numerical errors. 
To further improve the frequency resolution, we use linear prediction~\cite{White2008} extrapolating the correlations in momentum space up to times $2 t_{\rm max}$. We then apply a Gaussian windowing function ${\rm exp} \left[ -t^2/(2 t_{\rm max}^2) \right]$ to avoid ringing effects.

The strength of the backscattering interaction $g_{\rm bs}$ can be tuned in the lattice model by a second-neighbor exchange interaction $J_2$. We employ this fact to check the behavior of the dynamical correlations in the non-interacting limit, correpsonding to $J_{2,c}\simeq 0.24 J$. We note however that in the presence of DM interactions, tuning to the non-interacting limit requires simultaneously introducing a second-neighbor DM term $D_2$ whose strength is given by~\eqref{ap104} (see Appendix \ref{app:j1j2} for more details).
%$J_1-J_2$ chain with DM interactions is described in the Appendix \ref{app:j1j2}.

Below we discuss the numerical results for different orientations of the magnetic field with respect to the DM axis. In all cases, we observe excellent agreement with the analytical results obtained in the vicinity of $k=0$, as can be seen from the fits of the dispersions obtained numerically to the analytical form in each case. We note that while we observe some variations in the effective low energy velocity $v/J$ and dimensionless interaction strength $\delta$ depending on the orientation of the field, these could arise due to the finite, and not particularly small, value of DM interaction strength $D/J=0.3$ used in the simulations in order to achieve a better numerical accuracy.

\subsection{${\bf h}\parallel {\bf D}$}
\label{sec:h||d_dmrg}

\begin{widetext}

\begin{figure}[htb]
\centering 
	\begin{overpic}[width=0.32\columnwidth]{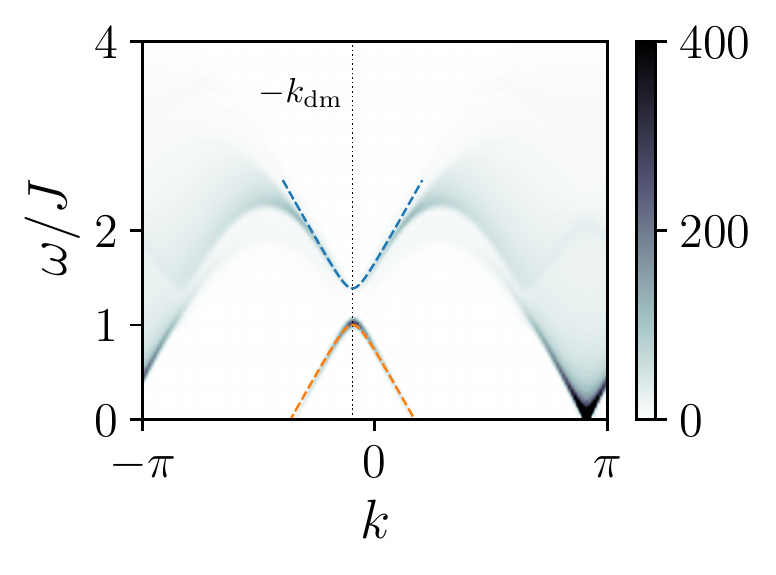} \put (0,67) {(a)} \end{overpic}
	\begin{overpic}[width=0.32\columnwidth]{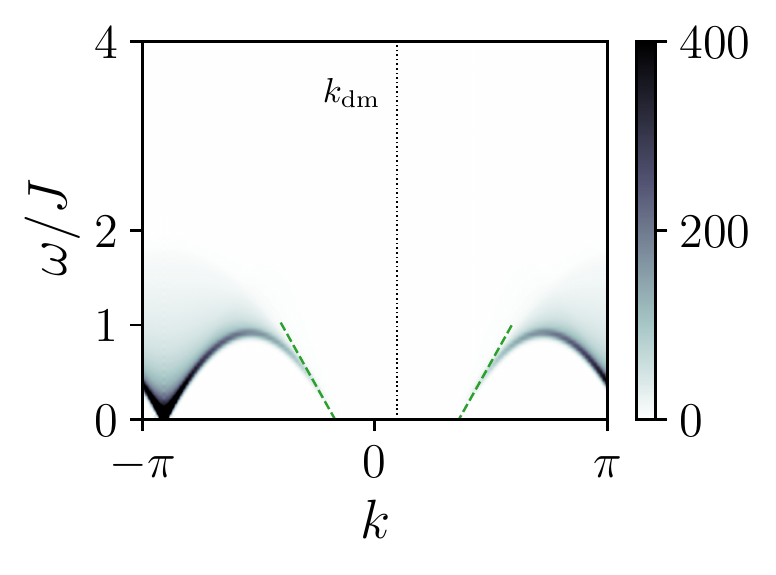} \put (0,67) {(b)} \end{overpic}
	\begin{overpic}[width=0.33\columnwidth]{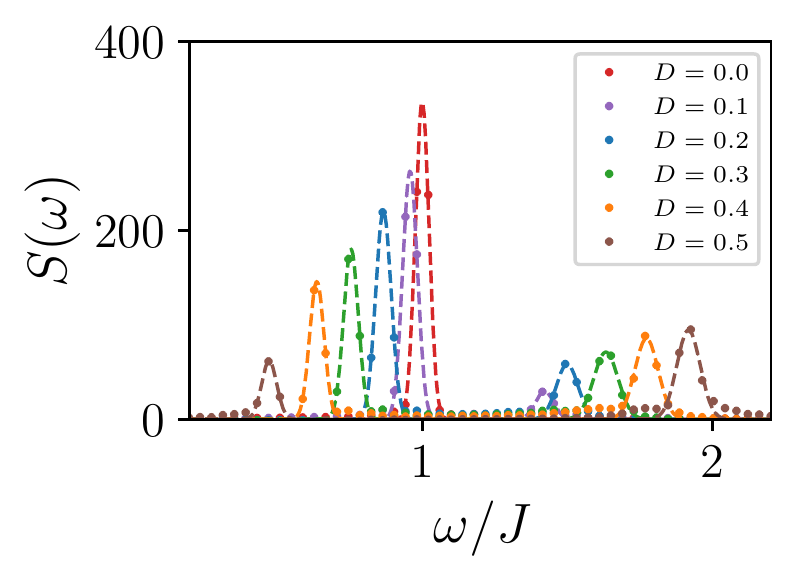} \put (0,65) {(c)} \end{overpic}
\caption{
Spectral functions obtained numerically for ${\bf h}\parallel {\bf D}$.
(a),(b) The transverse correlations $S^{+-}(k,\omega)$, and $S^{-+}(k,\omega)$, respectively, for $h/J=1,\ D/J=0.3$. Dashed lines indicate fits to the analytical dispersions $\omega_{\pm}(k,h)$ given by Eq.~\eqref{5.6} (and shown by blue and orange lines in Fig.~\ref{fig:1}(a)) in (a) and $\omega_{+}(-k,-h)$ (shown by green line in Fig.~\ref{fig:1}(a)) in (b) yielding $\delta=0.16, v/J=1.43$. (c) Cuts of  $S^{+-}(\omega)$ along $k=0$ for different values of $D$, showing a non-vanishing spectral weight of the two branches.
}
\label{fig:dmrg_par}
\end{figure}

\end{widetext}

When the magnetic field is parallel to the DM axis we observe that the dynamic structure factor $S^{+-}(k,\omega)$ is indeed boosted to momentum $k+k_{\rm dm}$ as expected from the discussion in Sec.~\ref{sec:1} and the detailed analysis in Sec.\ \ref{sec:h||d}. 
The structure factor $S^{-+}(k,\omega)$ is boosted in the opposite direction to $k-k_{\rm dm}$.  
This can be clearly seen in Figs.~\ref{fig:dmrg_par}(a) and~\ref{fig:dmrg_par}(b) respectively. Considering the dynamical correlations $S^{+-}(\omega)$ at $k=0$ one can now observe two peaks whose position and intensity depends on the strength of the DM interaction (see Fig.~\ref{fig:dmrg_par}(c)). According to \eqref{ou6} and \eqref{5.7}, this result should be compared with $A_\pm(k=0,h)$ in Fig.~\ref{fig:1}(b). And, indeed, the intensity $A_{+}(k=0,h)$ of the upper, magnetization-current-like mode is increasing as function of the DM parameter $D$ which enters \eqref{5.7} via $k_{\rm dm}$ given by \eqref{2.3}.

\subsection{${\bf h}\perp{\bf D}$}
\label{sec:DperpH_dmrg}

\begin{widetext}

\begin{figure}[htb]
\centering 
	\begin{overpic}[width=0.32\columnwidth]{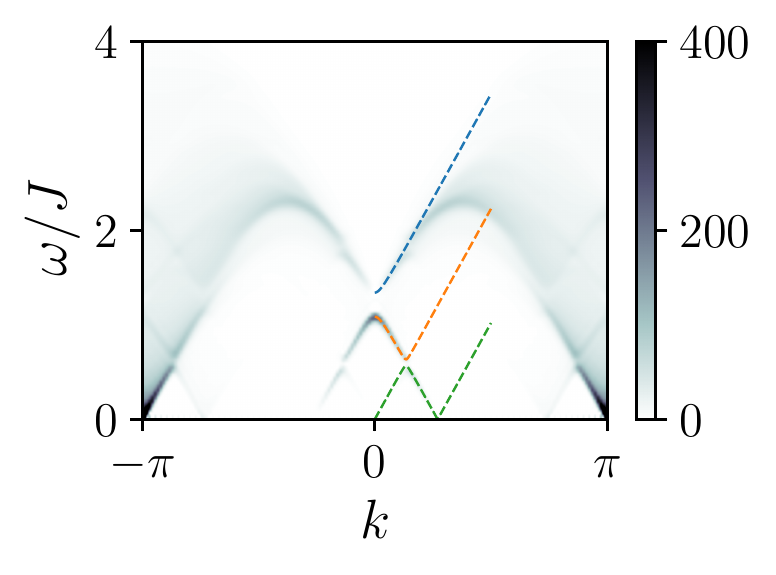} \put (0,67) {(a)} \end{overpic}
	\begin{overpic}[width=0.32\columnwidth]{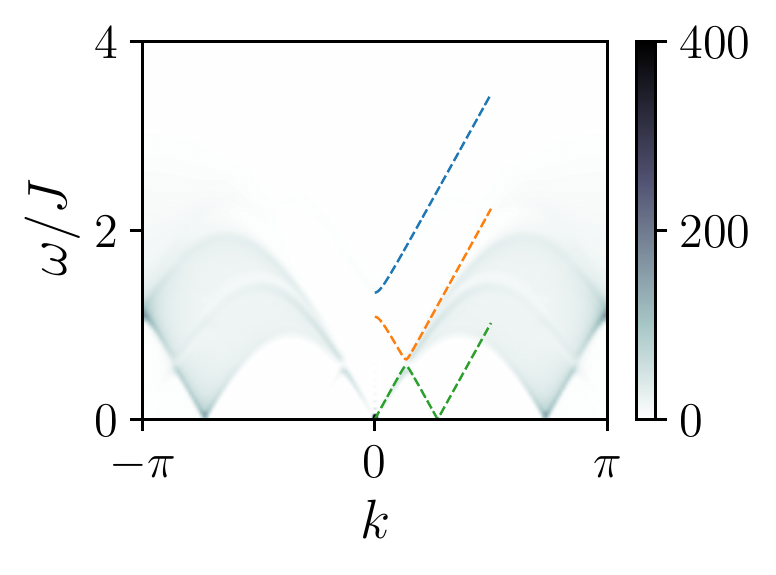} \put (0,67) {(b)} \end{overpic}
	\begin{overpic}[width=0.32\columnwidth]{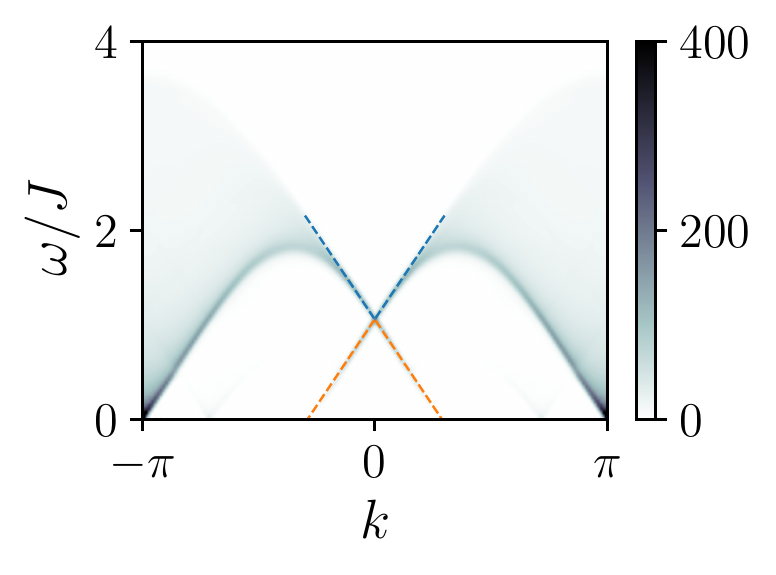} \put (0,67) {(c)} \end{overpic}
\caption{
Spectral functions obtained numerically for ${\bf h}\perp {\bf D}$ for $h/J=1,\ D/J=0.3$.
(a),(b) Transverse and longitudinal correlations $S^{+-}(k,\omega)$, and $S^{zz}(k,\omega)$, respectively. Dashed lines indicate fits to the analytical dispersions $\omega_{0,1,2}$ in~\eqref{A.2} with $\delta=0.12, v/J=1.43$. (c) Transverse correlations $S^{+-}(k,\omega)$ in presence of a second-neighbor exchange $J_2/J=0.24$ and second-neighbor DM $D_2$ given by~\eqref{ap104} 
corresponding to vanishing backscattering interactions. 
Dashed lines are fits to analytical dispersions \eqref{ap100} for non-interacting spinons, $\delta=0$, yielding $v/J=1.17$.}
\label{fig:dmrg_perp}
\end{figure}

\end{widetext}

Next we consider the case of magnetic field perpendicular to the DM axis. 
The transverse and longitudinal components of the dynamical susceptibility, which are now coupled, are shown in Fig.~\ref{fig:dmrg_perp}(a) and~\ref{fig:dmrg_perp}(b) respectively. These plots need to be compared with Figures \ref{fig:2} and \ref{fig:3} - and the agreement is excellent. Avoided crossing between $\omega_1(k)$ and $\omega_2(k)$ branches, predicted in Section ~\ref{sec:DperpH}, is very clearly visible in the numerical data.
The near invisibility of $\omega_{0,1}(k)$ branches in $S^{zz}(k,\omega)$, Fig.~\ref{fig:dmrg_perp}(b), is fully consistent with their very small spectral weights as shown in Fig.~\ref{fig:3}.

Tuning to the limit of vanishing backscattering interaction by including second-neighbor exchange term $J_2=0.24J$ and DM term $D_2$ given by \eqref{ap104}, we obtain transverse dynamical correlations shown in Fig.~\ref{fig:dmrg_perp}(c). As expected, in this case the gap at $k=0$ closes and we observe two linearly dispersing branches. The third, ``acoustic" branch $\omega_2(k)$ in \eqref{ap100}, is not visible due to its exceedingly small spectral weight, as illustrated by green lines in Fig.~\ref{fig:2}(b) (see also discussion following \eqref{b27}).

\subsection{Arbitrary angle $\theta$ between ${\bf h}$ and ${\bf D}$}
\label{sec:arbDH_dmrg}

\begin{figure}[htb]
\centering
	\begin{overpic}[width=\columnwidth]{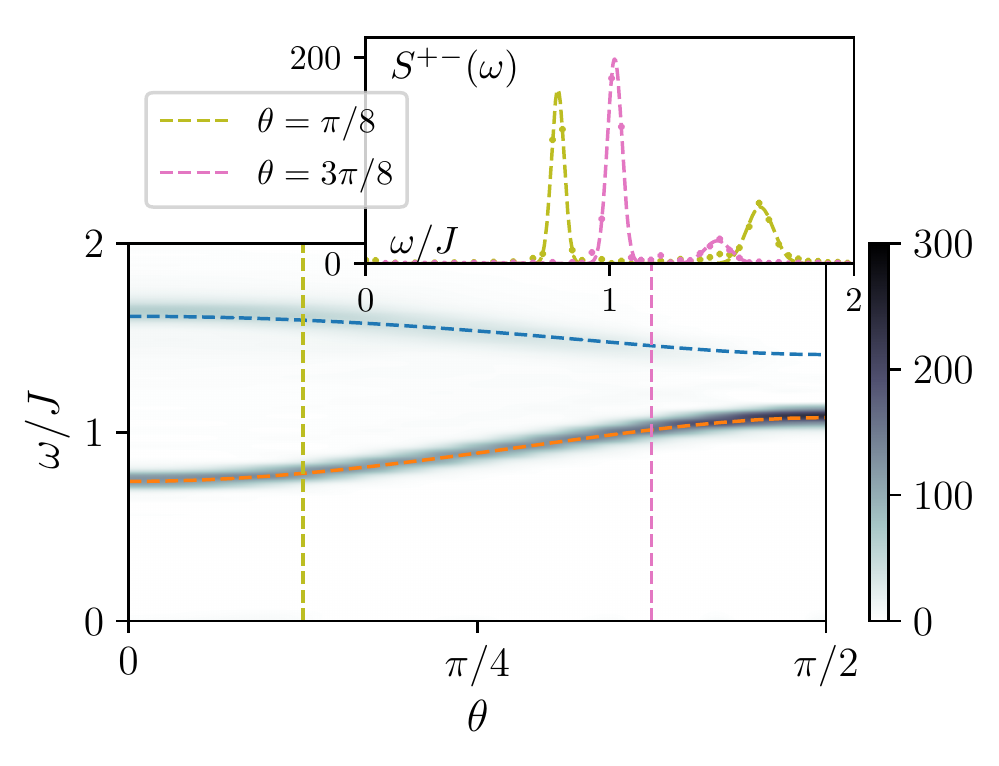} \put (0,70) {(a)} \end{overpic}
	\begin{overpic}[width=\columnwidth]{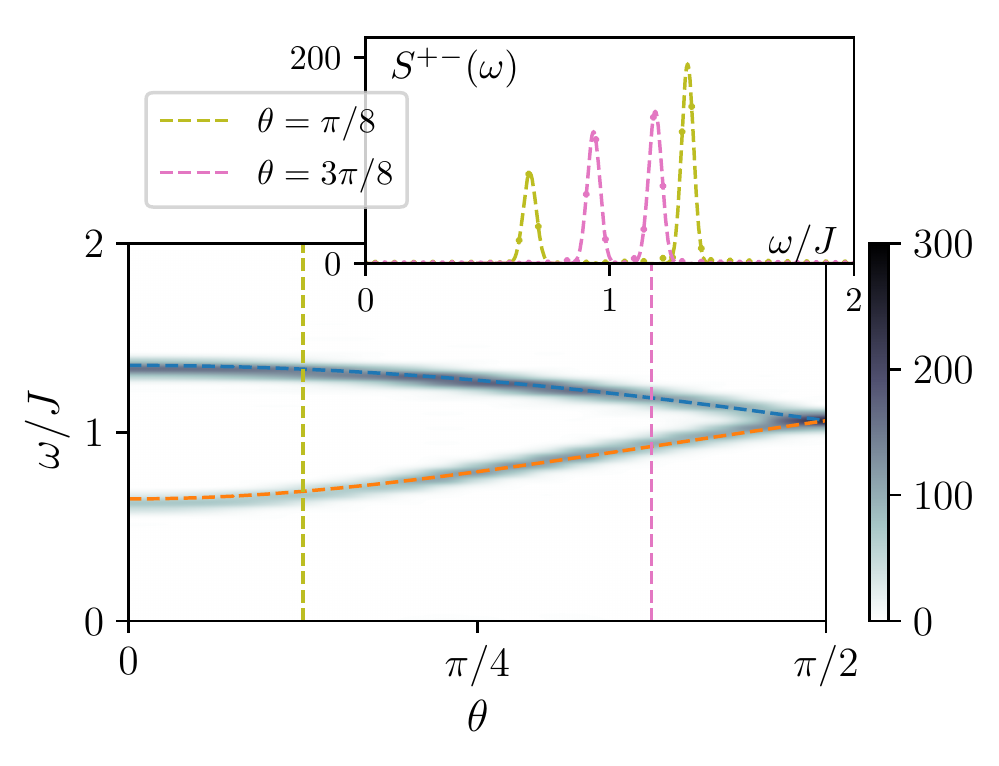} \put (0,70) {(b)} \end{overpic}
\caption{
Spectral functions $S^{+-}(\theta,\omega)$ for $k=0$ obtained numerically for $h/J=1,\ D/J=0.3$.
(a) $J_2=D_2=0$, (b) $J_2/J=0.24$ and $D_2$ given by \eqref{ap104}. 
Dashed lines are fits to $\Omega_{\pm}$ in~\eqref{5.15} yielding $\delta=0.15,\ v/J=1.35$ in (a), and $\delta=0,\ v/J=1.18$ in (b). The relative intensity of the two branches as the angle is varied can be seen in the insets. A qualitative agreement with the spectral weights obtained analytically and shown in Fig.~\ref{fig:5}(a) is clearly observed.}
\label{fig:dmrg_arbDH}
\end{figure}

Finally, we consider the case of an arbitrary angle $\theta$ between the field and the DM axis, focusing on the response at $k=0$. Transverse correlations $S^{+-}(\theta,\omega)$ are shown in Fig.~\ref{fig:dmrg_arbDH} both in the Heisenberg limit ($J_2=0$) and the limit of vanishing backscatterinig ($\delta=0$). The data is in agreement with analytical analysis in Section \ref{sec:arbDH} 
regarding both the angular dependence of the excitation energies, \eqref{5.15} and Fig.~\ref{fig:4}, and the intensities, Fig.~\ref{fig:5}(a). Note, for example, that while for strongly interacting spinons the intensity of the upper mode is much smaller than that for the lower mode, Fig.~\ref{fig:dmrg_arbDH}(a), for the non-interacting ones, Fig.~\ref{fig:dmrg_arbDH}(b), the situation is somewhat reversed. This is also present in Fig.~\ref{fig:5}(a) where dotted blue line, corresponding to the intensity of $\Omega_{+}(\theta)$ for $\delta=0$ lies above the dotted orange one for the intensity of $\Omega_{-}(\theta)$.

Once again, the analytical hydrodynamic approximation captures (and explains) all essential features of the spin chain response at small momentum.

\section{Discussion}
\label{sec:conclude}

Majority of spectral weight in Figures~\ref{fig:dmrg_par} and  ~\ref{fig:dmrg_perp} is contained in spinon continua that become very pronounced {\em away} from $k \approx 0$ regime on which we focus in this work.
Theoretical description of these continua is well established within the standard framework of bosonization \cite{Gogolin1998} as well as non-linear bosonization corrections to it \cite{Imambekov2012,Karimi2011,Sirker2011}.
At this point we only note that faint but visible low-energy spectral weight near $k=\pm\pi \mp B/v$ visible in Fig.~\ref{fig:dmrg_perp}(a) is the contribution of the staggered dimerization operator which admixes to the transverse spin response in this low-symmetry geometry, see \cite{Chan2017} for more details (notice that magnetic field is oriented along the $\hat{x}$ axis there). Much bigger spectral weight at the same $k=\pm\pi \mp B/v$ in Fig.~\ref{fig:dmrg_perp}(b) is the standard longitudinal spin contribution from $N^z$ component of the Ne\'el operator \cite{Gogolin1998}.

In the region of our interest $k \approx 0$, however, the spectral lines are very narrow and very well approximated by delta-function peaks as predicted by our hydrodynamic theory. This feature has 
to do with the linear dispersion of Dirac fermions that underline our low-energy Hamiltonian \eqref{3.1}. Deviations of the dispersion from the strictly linear form, that become important away from zero momentum,
will give spectral lines finite width even at small $k$ \cite{Imambekov2012} - this interesting theoretical problem is outside the scope of the current study.

Symmetry-breaking DM interaction, that provides magnetization-current-like branches of spin excitations with finite intensity at $k=0$, is also responsible for the finite linewidth of the ESR spectra, as described in \cite{Oshikawa2002,Furuya2017}. Our linearized hydrodynamic approximation does not account for the self-energy corrections of that kind. For completeness, we mention also that in a quantum wire setting the linewidth mostly comes from the coupling to gapless charge degrees of freedom  \cite{Pokrovsky2013,Pokrovsky2017},

To summarize, the presented hydrodynamic approach captures all essential features of the nearly uniform, i.e.\ $k \approx 0$, dynamic spin response of the Heisenberg chain perturbed by the uniform DM interaction. The described approach is simple, internally consistent, and provides an intriguing connection of this interacting spinon liquid picture with existing literature on spin dynamics of neutral Fermi liquids \cite{Leggett1970}. Our theory is supported by extensive comparison with numerical MPS-based simulations reported here. Its key predictions for the ESR experiments have been successfully verified very recently \cite{Povarov2022}. Experimental verification of the avoided level crossing, such as reported in Figures \ref{fig:2}(a) and \ref{fig:dmrg_perp}(a,b), which requires inelastic neutron scattering measurements, is highly desirable. 

\section*{Acknowledgements}

We thank Leon Balents for insightful discussions of the spin chain hydrodynamics and Kirill Povarov, Timofei Soldatov, and Alexander Smirnov for discussions of ESR experiments in quasi-one-dimensional antiferromagnet K$_2$CuSO$_4$Br$_2$.  
R.B.W.\ and O.A.S.\ were supported by the NSF CMMT program under Grant No.\ DMR-1928919.
A.K.\ acknowledges funding by the Israeli Council for Higher Education support program for hiring outstanding faculty members in quantum science and technology in research universities.

\appendix
\section{Derivation of Heisenberg equations of motion for the chiral spin currents \eqref{4.1}}
\label{app:A}
In order to derive the Heisenberg equations of motion for the chiral spin currents \eqref{4.1}, we need to compute the commutator $i[\hat H_0+\hat H_\mathrm{ bs}+\hat V,\hat{J}_{r}^a(x)]$.
Here $r=R = +1$ and $r=L = -1$.
Calculate $i[\hat H_0,\hat{J}_{r}^a(x)]$ first,
\begin{align}
&i[\hat H_0,\hat{J}_{r}^a(x)]
=i[\frac{2\pi v}{3}\int \mathrm dx':\hat {J}_{r'}^b(x')\hat {J}_{r'}^b(x'):,\hat{J}_{r}^a(x)] \nonumber\\ %\label{A1}\\
&\begin{aligned}
&=i\frac{2\pi v}{3}\int \mathrm dx'\lim_{\epsilon\to0}
\{\hat {J}_{r'}^b(x'+\frac{\epsilon}{2})[\hat {J}_{r'}^b(x'-\frac{\epsilon}{2}),\hat{J}_{r}^a(x)]\\
&+[\hat {J}_{r'}^b(x'+\frac{\epsilon}{2}),\hat{J}_{r}^a(x)]\hat {J}_{r'}^b(x'-\frac{\epsilon}{2})\},
\end{aligned}\label{A2}
\end{align}
where we have used the definition of point-splitting to resolve the singularity of the product $\hat {J}_{r'}^b(x')\hat {J}_{r'}^b(x')$ at the same point $x'$.
Then we use the Kac-Moody algebra \eqref{eq1}  and find
\begin{align}
\begin{aligned}
&i[\hat H_0,\hat{J}_{r}^a(x)]
=i\frac{2\pi v}{3}\int \mathrm dx'
\lim_{\epsilon\to0}\{\hat {J}_{r'}^b(x'+\frac{\epsilon}{2})
i\delta_{r'r}\\
&[\frac{-r'}{4\pi} \delta'(x'-\frac{\epsilon}{2}-x)\delta^{ba} 
+ \delta(x'-\frac{\epsilon}{2}-x)  \epsilon^{bac} J_{r}^c(x'-\frac{\epsilon}{2})]\\
&+i\delta_{r'r}[\frac{-r'}{4\pi} \delta'(x'+\frac{\epsilon}{2}-x)\delta^{ba} \\
&+ \delta(x'+\frac{\epsilon}{2}-x)  \epsilon^{bac} J_{r}^c(x'+\frac{\epsilon}{2})]\hat {J}_{r'}^b(x'-\frac{\epsilon}{2})\}.
\end{aligned}
\end{align}
Finally, we use the operator product expansion \cite{Ludwig1995,Gogolin1998,Wang2020}
\begin{align*}
J_{r}^a(x)J_{r'}^b(x')
&=\frac{-\delta_{rr'}\delta^{ab}}{8\pi^2(x-x')^2}
+\frac{-r\delta_{rr'}\epsilon^{abc}}{2\pi(x-x')}J_{r}^c(x'),
\end{align*}
where $x-x'\to0^+$, to evaluate products of $\hat {J}_{r'}^b(x'+\frac{\epsilon}{2})J_{r}^c(x'-\frac{\epsilon}{2})$ and $J_{r}^c(x'+\frac{\epsilon}{2})\hat {J}_{r'}^b(x'-\frac{\epsilon}{2})$, and find that
\begin{align}
&\begin{aligned}
&i[\hat H_0,\hat{J}_{r}^a(x)]
=-\frac{2\pi v}{3}
\lim_{\epsilon\to0}\{
[\frac{r}{4\pi} \partial_x \hat {J}_{r}^a(x+\epsilon)\\
&+ \epsilon^{bac}\hat {J}_{r}^b(x+\epsilon) J_{r}^c(x)]\\
&+[\frac{r}{4\pi} \partial_x \hat {J}_{r}^a(x-\epsilon)
+ \epsilon^{bac} J_{r}^c(x)\hat {J}_{r}^b(x-\epsilon)]\}
\end{aligned}\\
&\begin{aligned}
&=-\frac{2\pi v}{3}
\{\frac{r}{2\pi} \partial_x \hat {J}_{r}^a(x)\\
&+\lim_{\epsilon\to0} \epsilon^{bac}
[\frac{-r\epsilon^{bcd}}{2\pi\epsilon}J_{r}^d(x)
+\frac{-r\epsilon^{cbd}}{2\pi\epsilon}J_{r}^d(x-\epsilon)]\}
\end{aligned}\\
&=-\frac{r v}{3} \Big(\partial_x\hat{J}_{r}^a(x) + \frac{1}{\epsilon} \epsilon^{abc} \epsilon^{dbc}[\hat{J}_{r}^d(x) - \hat{J}_{r}^d(x-\epsilon)]\Big)\nonumber\\
&=-rv\partial_x\hat{J}_{r}^a(x),\label{A6}
\end{align}
where we used $\epsilon^{abc} \epsilon^{dbc} = 2\delta^{ad}$.
The other two terms in the commutator are simpler to evaluate,
\begin{align}
&i[\hat H_\mathrm{bs},\hat{\mathbf J}_{r}(x)]
=rg_\mathrm{bs}[\frac{1}{4\pi}\partial_x\hat{\mathbf J}_{\bar r}(x)+\hat{ \mathbf J}_R(x)\times \hat{ \mathbf J}_L(x)],\label{A7}\\
&i[\hat V,\hat{\mathbf J}_{r}(x)]
=-(\mathbf h+r\tilde{\mathbf D})\times \hat {\mathbf J}_r(x).\label{A8}
\end{align}
Eq.\ \eqref{4.1} follows from \eqref{A6}-\eqref{A8}.

\section{The transverse  and longitudinal susceptibilities for  the case ${\bf H}\perp{\bf D}$ at finite $k$}
\label{app:a}

The  characteristic equation \eqref{5.9} gives an even sextic equation
\begin{align}
\omega^6+d_2\omega^4+d_1\omega^2+d_0=0,
\label{A.1}
\end{align}
where 
\begin{align}
-d_2
&=3(1-\delta^2)(vk)^2+2\frac{1+\delta^2}{(1-\delta)^2}h^2+2\frac{1-\delta}{1+\delta}\tilde{D}^2, \\
d_1
&=3[(1-\delta^2)(vk)^2]^2+[\frac{1+\delta^2}{(1-\delta)^2}h^2+\frac{1-\delta}{1+\delta}\tilde{D}^2]^2\nonumber\\
&+4(\frac{\delta h}{1-\delta})^2[(1-\delta^2)(vk)^2-(\frac{h}{1-\delta})^2], 
\end{align}
and
\begin{align}
-d_0=(1-\delta^2)(vk)^2[(1-\delta^2)(vk)^2-(\frac{1+\delta}{1-\delta}h^2+\frac{1-\delta}{1+\delta}\tilde{D}^2)]^2.
\end{align}
Since \eqref{A.1} is an cubic equation of $\omega^2$, the solutions can be constructed from the Viète's formula,
\begin{align}
\begin{aligned}
\omega_i^2(k)+\frac{d_2}{3}
=&2\sqrt{-\frac{\alpha_1}{3}}
\cos[\frac{1}{3}\cos^{-1}(\frac{3\alpha_0}{2\alpha_1}\sqrt{\frac{-3}{\alpha_1}})-\frac{2\pi}{3}i],
\end{aligned}\label{A.2}
\end{align}
where $i=0,1,2,$
$
\alpha_1
=(3d_1-d_2^{2})/3,
$
$\alpha_0
=(2d_2^{3}-9d_2d_1+27d_0)/27.
$
From \eqref{5.2}, the analytical form of the longitudinal and transverse dynamical retarded susceptibilities can be expressed as
\begin{align}
\chi^{+-}(k,\omega)
=&\chi_0\sum_{i=0}^2\sum_{\eta=\pm}
\frac{A_{i\eta}^{+-}(k)}{\omega-\eta \omega_i(k)+i0^+},\label{A.3}\\
\chi^{-+}(k,\omega)|_h=&\chi^{+-}(-k,\omega)|_{-h},\label{A.4}\\
\chi^{zz}(k,\omega)
=&\chi_0\sum_{i=0}^2\sum_{\eta=\pm}
\frac{A_{i+}^{zz}(k)}{\omega-\eta \omega_i(k)+i0^+},\label{A.5}
\end{align}
where \eqref{A.4} is the Onsager relation and the spectral weights are given by
\begin{align}
A_{j \eta}^{+-}(k)
&=\frac{\eta a_0+a_1\omega_j + \eta a_2\omega_j^2 + a_3\omega_j^3 + \eta a_4\omega_j^4 + a_5\omega_j^5}{2\omega_j(\omega_j^2-\omega_{j+1}^2)(\omega_j^2-\omega_{j+2}^2)},\label{A.6}\\
A_{j\eta}^{zz}(k)
&=\eta\frac{b_0+ b_2\omega_j^2+ b_4\omega_j^4}{2\omega_j(\omega_j^2-\omega_{j+1}^2)(\omega_j^2-\omega_{j+2}^2)} , \label{A.7}
%=-A_{0-}^{zz}(k),\label{A.7}
\end{align}
for $j=(0,1,2)$. Here indices $j+1,j+2$ are a short-hand notation for $j+1, j+2~{\rm mod}(3)$ and $\eta=\pm$. The coefficients are  
\begin{align}
a_0
&=2 (1 + \delta) (k v)^2 [(1 - \delta^2) (k v)^2-(\frac{1-\delta}{1+\delta}\tilde{D} ^2 +\frac{1+\delta}{1-\delta} h^2)]^2, \\
a_1
&=-2(1 + \delta)^3 h (kv)^2  [(kv)^2 - (\frac{\tilde{D}}{1+\delta})^2+(\frac{h}{1-\delta})^2], \\
a_2
&=-(1 -\delta^2) (1 +\delta) [4  (vk)^4 +(\frac{\tilde{D}}{1+\delta})^4+2(\frac{h}{1-\delta})^4\nonumber\\
&+ (1+\delta)(-3+(2-3\delta)\delta)(\frac{\tilde{D}}{1+\delta})^2(\frac{h}{1-\delta})^2\nonumber\\
&+ (1 - \delta^2) (vk)^2 [-(\frac{\tilde{D}^2}{1+\delta})+2(1+3\delta)(\frac{h}{1-\delta})^2], \\
a_3
&=2\frac{1+\delta}{1-\delta} h[2\delta (v k)^2-((1-\delta)(\frac{\tilde{D}}{1+\delta})^2+(1+\delta)(\frac{h}{1-\delta})^2],\\
a_4
&= 2 (1 + \delta)(vk)^2 +\frac{2 h^2}{1-\delta}+\frac{\tilde{D}^2}{1+\delta} ,\\
a_5
&=\frac{2h}{1-\delta},
\end{align}
\begin{align}
b_0
&= (1 + \delta)(vk)^2 [(1 - \delta^2)(vk)^2 -(\frac{1-\delta}{1+\delta}\tilde{D}^2+ \frac{1+\delta}{1-\delta} h^2 )]^2,\\
b_2
&=(1 - \delta) (-\frac{\tilde{D}^4}{(1+\delta)^2}- 2 (vk)^4(1 + \delta)^2 \nonumber\\
&- \frac{(1+\delta)(1+\delta^2)2 h^2 (vk)^2}{(1-\delta)^3}  +\tilde{D}^2[(vk)^2 -\frac{h^2}{1-\delta^2}]), 
\end{align}
and 
\begin{align}
b_4
=\frac{\tilde{D}^2}{1+\delta}  + (1 + \delta)(vk)^2 .
\end{align}

The limit of non-interacting spinons, $\delta =0$, leads to dramatic simplifications. 
We note that for $\delta =0$ coefficients of \eqref{A.1} depend only on $vk$ and the total field $\Xi \equiv \sqrt{h^2 + \tilde{D}^2}$, 
which is $\delta=0$ version of the field $B$ in \eqref{jan1}.
Moreover, 
substitution $y=\omega^2 - v^2 k^2$ reduces \eqref{A.1} to the very simple factorized form
\be
y \Big( (y-\Xi^2)^2 - 4 \Xi^2 v^2 k^2\Big) = 0,
\label{ap99}
\ee
from which we find non-interacting analogues of \eqref{5.14}-\eqref{o28}
\bea
&&\omega_0(k) = \Xi + v |k|, \quad
\omega_1(k) = \Xi - v |k|, 
\nonumber\\
&&\omega_2(k) = v |k| .
\label{ap100}
\eea
Here $\tilde{D} = v D/J$, see \eqref{5.8}. 
Dispersions \eqref{ap100} are shown by dashed lines in Fig.\ \ref{fig:2}(a). Spectral weights associated with these modes of non-interacting spinons are (see \eqref{ap101})
\begin{align}
A^{+-}_{0\pm}
&=\pm\frac{\Xi+vk}{4}(1\pm\frac{h}{\Xi})^2,\label{b22}\\
A^{+-}_{1\pm}
&=\pm\frac{\Xi-vk}{4}(1\pm\frac{h}{\Xi})^2,\label{b23}\\
A^{+-}_{2\pm}
&=\pm\frac{1}{2}(\frac{\tilde{D}}{\Xi})^2vk,\label{b24}\\
A^{zz}_{0\pm}
&=\pm\frac{\Xi+vk}{4}(\frac{\tilde{D}}{\Xi})^2,\\
A^{zz}_{1\pm}
&=\pm\frac{\Xi-vk}{4}(\frac{\tilde{D}}{\Xi})^2,\\
A^{zz}_{2\pm}
&=\pm\frac{1}{2}(\frac{h}{\Xi})^2vk.\label{b27}
\end{align}
Our DMRG data on the ``non-interacting" chain, Figure \ref{fig:dmrg_perp}(c), agree with these analytical results very well. The ``optical" branches $\omega_{0,1}(k)$ with highly linear dispersion are very well resolved, in agreement with \eqref{ap100}, \eqref{b22}, and \eqref{b23}. 
The absence of the ``acoustic" branch $\omega_{2}(k)$ in Fig.\ \ref{fig:dmrg_perp}(c) is naturally explained by the smallness of the 
spectral weight \eqref{b24} in $D/\Xi$ ratio as well as its linear in $k$ form.
All these features are also clearly illustrated by our Figure \ref{fig:2}(b), where both non-interacting (dashed lines) and interacting ($\delta=0.12$, solid lines) spectral weights are plotted.

\section{Dynamic susceptibilities at $k=0$ for the general case of the angle $\theta$ between ${\bf H}$ and ${\bf D}$ }
\label{app:b}

The analytical form of the longitudinal and transverse dynamical retarded susceptibilities at $k=0$ for  the case ${\bf H}$ in arbitrary directions with ${\bf D}$ can be expressed as
\begin{align}
\chi^{+-}(\omega,\theta)
=&\chi_0\sum_{\mu=\pm}\sum_{\eta=\pm}\frac{\tilde A_{\mu \eta}^{+-}(\theta)}{\omega-\eta \Omega_\mu(\theta)+i0^+},\\
\chi^{-+}(\omega,\theta)|_{h}
=&\chi^{+-}(\omega,\theta)|_{-h},\label{B.1}\\
\chi^{zz}(\omega,\theta)
=&\chi_0\sum_{\mu=\pm}\sum_{\eta=\pm}
\frac{\tilde A_{\mu \eta}^{zz}(\theta)}{\omega-\eta \Omega_\mu(\theta)+i0^+},
\label{B.2}
\end{align}
where \eqref{B.1} is the Onsager relation, and the spectral weights are ($\mu = \pm, \eta = \pm$)
\begin{align}
\tilde A_{\mu \eta}^{+-}(\theta)
&=\frac{\eta \tilde a_0 + \tilde a_1\Omega_\mu + \eta \tilde a_2\Omega_\mu ^2 + \tilde a_3\Omega_\mu^3}{2\Omega_\mu(\Omega_\mu^2-\Omega_{-\mu}^2)},\label{B.3}\\
\tilde A_{\mu \eta}^{zz}(\theta)
&=\eta \frac{(\sin\theta\tilde{D})^2}{1+\delta}\frac{\tilde  b_0+ \Omega_\mu^2}{2\Omega_\mu(\Omega_\mu^2-\Omega_{-\mu}^2)},\label{B.4}
\end{align}
%and cyclic permutation for the other ``$-$" component 
with 
\begin{align}
\tilde a_0
&=-2\frac{(1+\delta)^2}{(1-\delta)^3}h^4-\frac{(1-\delta)(3+2\cos2\theta)}{2(1+\delta)^2}\tilde{D}^4\nonumber\\
&+\frac{1+7\cos2\theta+4\sin^2\theta\delta-(7+\cos2\theta)\delta^2}{2(1-\delta^2)(1-\delta)}h^2\tilde{D}^2, \\
\tilde a_1
&=\frac{-2h}{1-\delta} [(\frac{1+\delta}{1-\delta}h)^2 - \frac{\delta+\cos2\theta}{1+\delta}\tilde{D}^2], \\
\tilde a_2
&=\frac{2h^2}{1 -\delta}+ \frac{3+\cos2\theta}{2(1 +\delta) }\tilde{D}^2, \\
\tilde a_3
&=\frac{2h}{1 -\delta}, 
\end{align}
and 
\begin{align}
-\tilde b_0
= h^2+\frac{1-\delta}{1+\delta}\tilde{D}^2.
\end{align}

\begin{figure}[hbt]
\centering
\includegraphics[width=.45\linewidth]{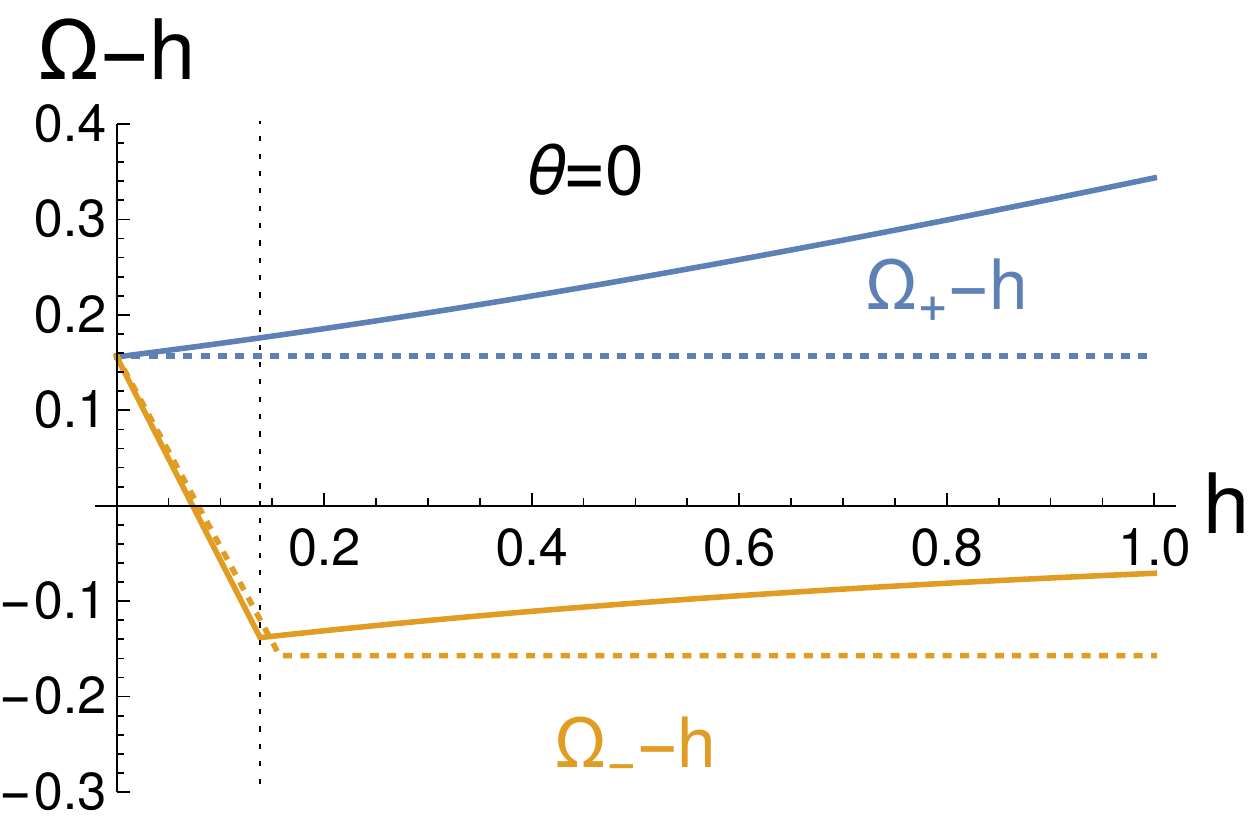}
\includegraphics[width=.45\linewidth]{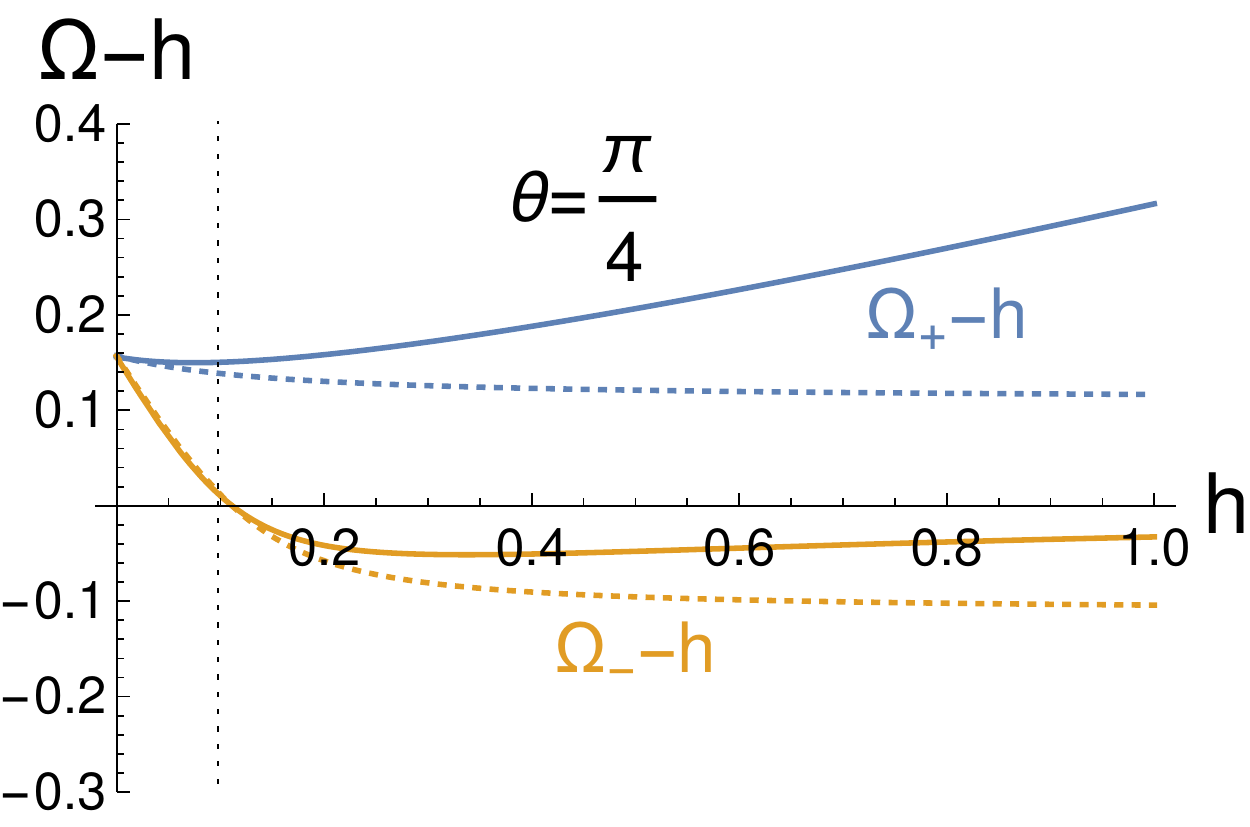}
\includegraphics[width=.45\linewidth]{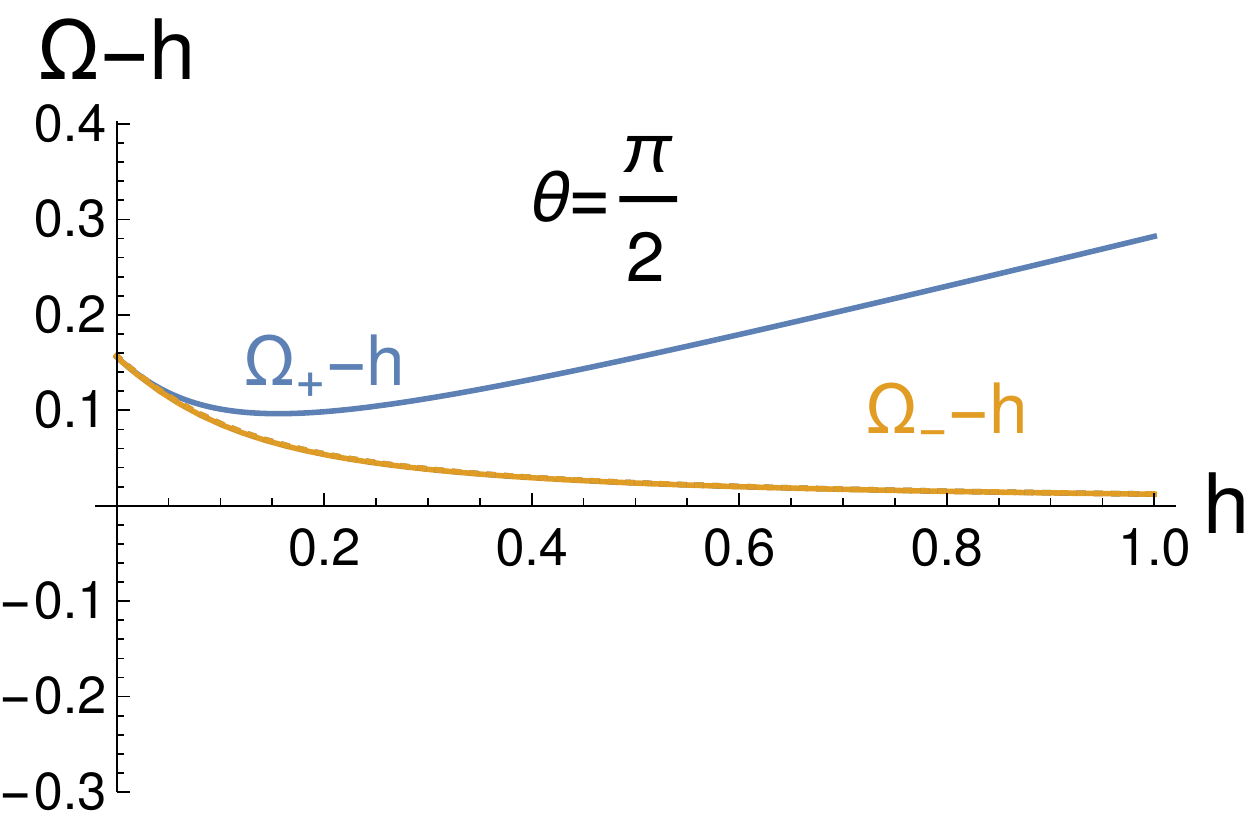}
\caption{
The reduced frequencies $\Omega_+(\theta)-h$ (blue line) and $\Omega_-(\theta)-h$ (orange line)  as a function of the field $h$ for $\theta=0$, $\frac{\pi}{4}$, and $\frac{\pi}{2}$. 
Solid lines are for $\delta=0.12$ and dotted lines are for $\delta= 0$.
Dotted vertical line shows the crossover field $h_c=\frac{1-\delta}{1+\delta}\tilde{D}\cos\theta
\approx(1-\delta)\frac{\pi }{2}D\cos\theta$.}
\label{fig:7}
\end{figure}

\begin{figure}[hbt]
\includegraphics[width=.45\linewidth]{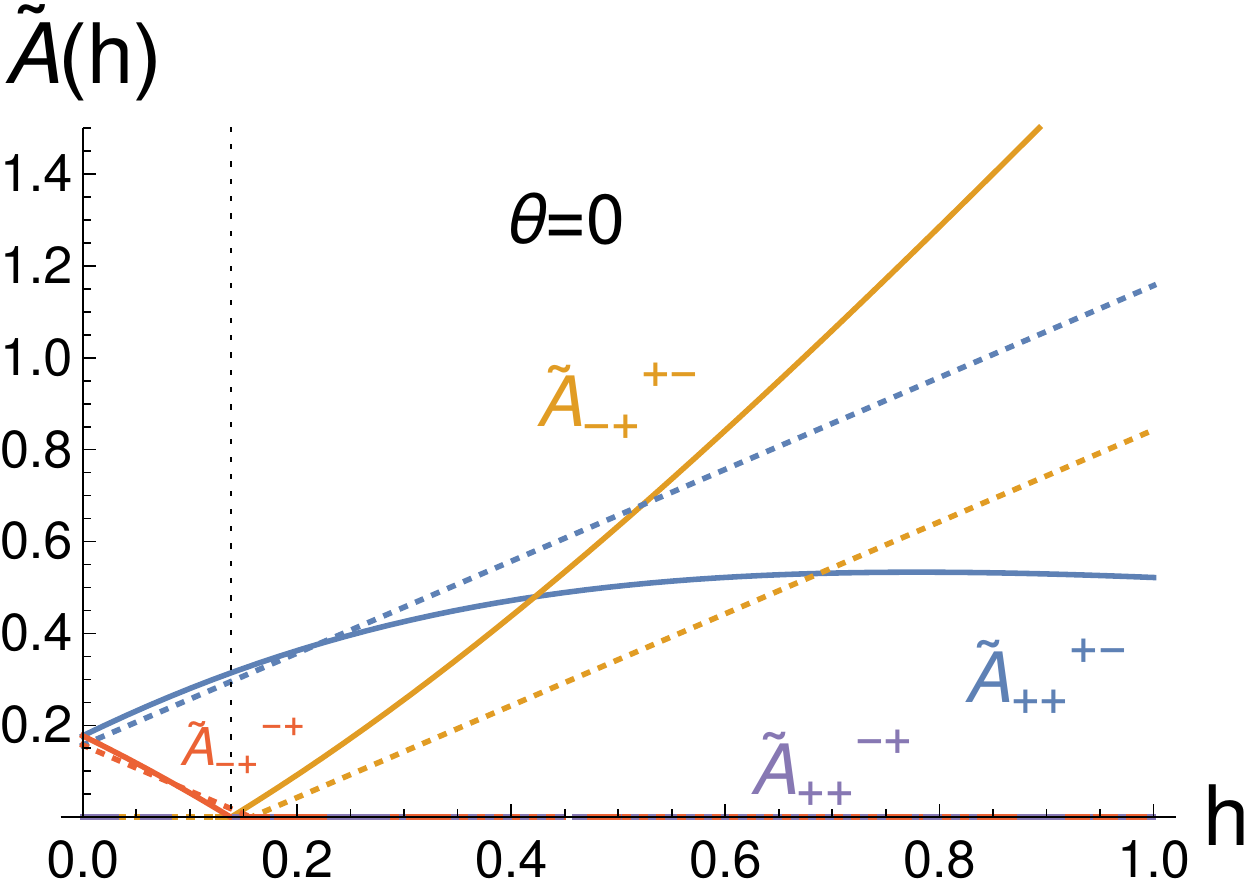}
\includegraphics[width=.45\linewidth]{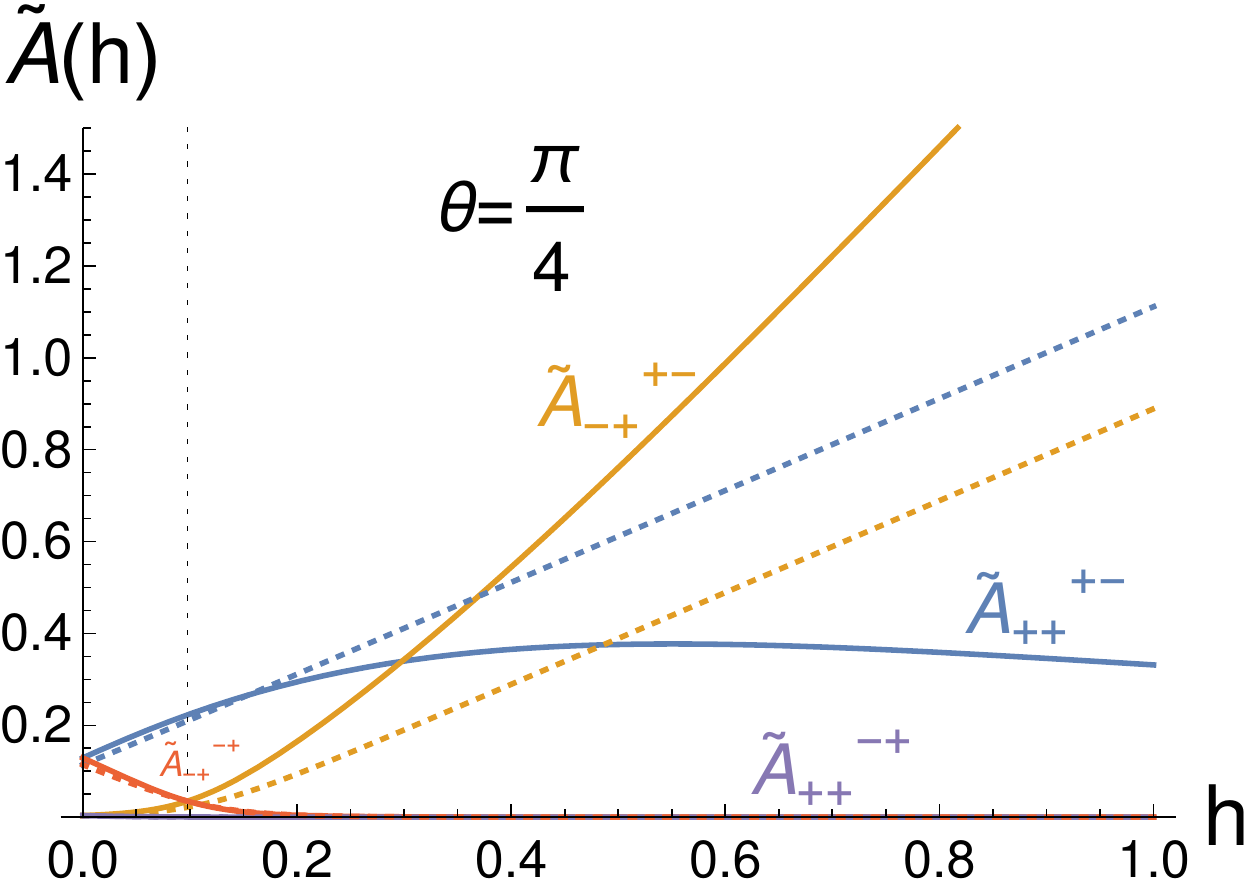}
\includegraphics[width=.45\linewidth]{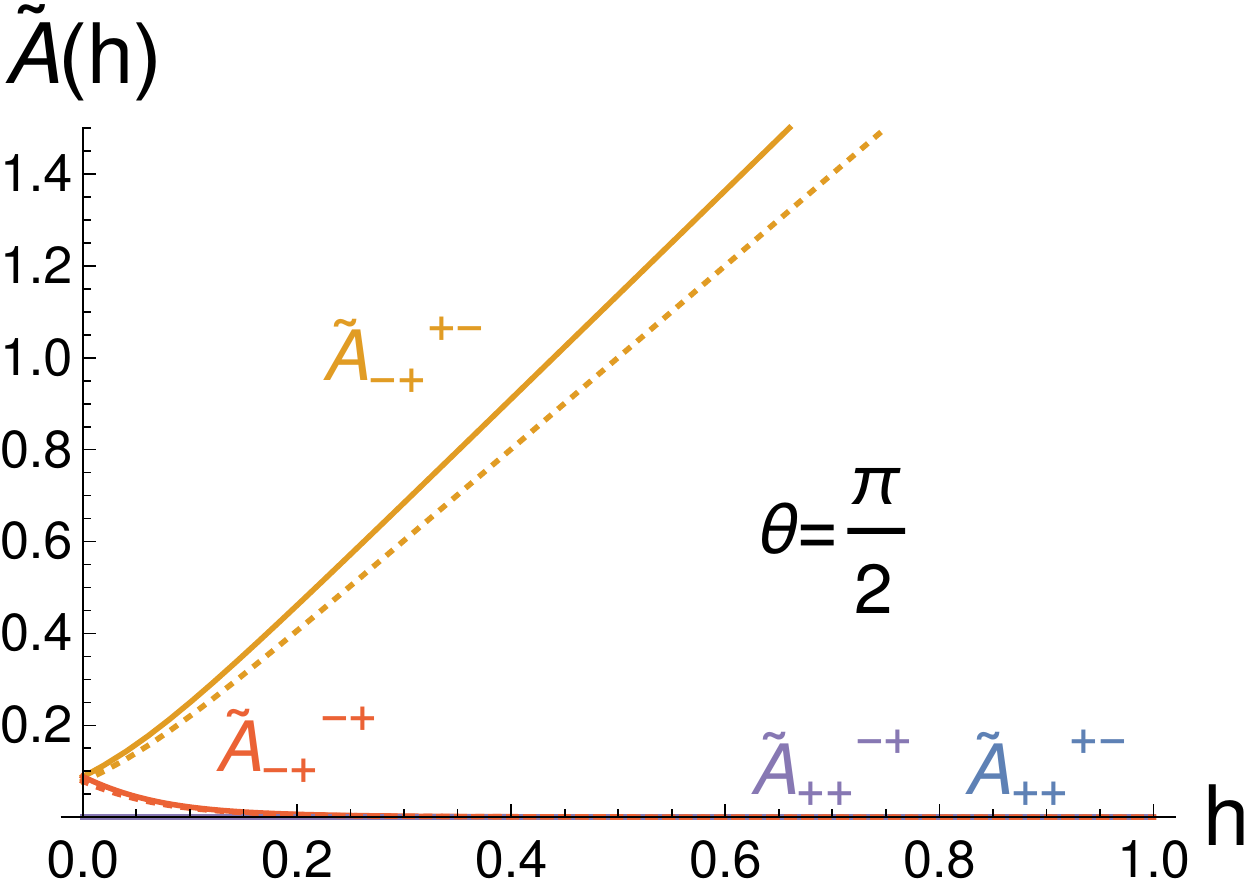}
\caption{The spectral weights $\tilde{A}_{+ +}^{+-}$ (blue line),  $\tilde{A}_{- +}^{+-}$ (orange line) $\tilde{A}_{+ +}^{-+}=0$ (purple line),  $\tilde{A}_{- +}^{-+}$ (red line) as a function of the field $h$ for $\theta=0$, $\frac{\pi}{4}$, and $\frac{\pi}{2}$. 
Solid lines are for $\delta=0.12$ and dotted lines are for $\delta= 0$.
Dotted vertical line shows the crossover field $h_c=\frac{1-\delta}{1+\delta}\tilde{D}\cos\theta
\approx(1-\delta)\frac{\pi }{2}D\cos\theta$.}
\label{fig:8}
\end{figure}

We plot the reduced frequency $\Omega_\pm(\theta)-h$ as a function of the field $h$ for $\theta=0$, $\frac{\pi}{4}$, and $\frac{\pi}{2}$ in Fig.\ \ref{fig:7}, corresponding spectral weights are plotted in Fig.\ \ref{fig:8}.
For $\theta=0$, there is a kink in the lower branch, at 
$h_c=\frac{1-\delta}{1+\delta}\tilde{D} \approx(1-\delta)\frac{\pi }{2}D$.  This is explained \cite{Povarov2022} by the `switching' of the contribution from $\chi^{-+}$ for $h < h_c$ to that 
from $\chi^{+-}$ for $h > h_c$. This is also seen from \eqref{5.15} since at $\theta = 0$ the $\mu = -1$ expression reads 
$\Omega_{-}(0) = | \frac{h}{1-\delta}  - \sqrt{\big(\frac{h \delta}{1-\delta}\big)^2 + \frac{1-\delta }{1+\delta }\tilde{D}^2}|$. The kink happens when the argument of the absolute value changes sign.

For any $\theta\neq0$ the lower branch smoothens out, as \eqref{5.15} predicts, although the minimum at $h_c \approx(1-\delta)\frac{\pi }{2}D\cos(\theta)$ can still be observed for $\theta \leq \pi/4$.
For $h\leq h_c$, the spectral weight of the lower mode of $\chi^{-+}$, $\tilde{A}_{-+}^{-+}$, is always finite and bigger than that for $\chi^{+-}$, except for $\theta=\pi/2$.
Hence $h_c$ is also the crossover magnetic field such that the spectral weight of the lower mode for $\chi^{+-}$ starts to exceed that of $\chi^{-+}$.

\section{$J_1$-$J_2$ chain with DM interactions}
\label{app:j1j2}

Backscattering interaction $g_{\rm bs}$ in \eqref{3.3}, more specifically, its bare (or, initial) value, is a function of exchange interactions $J_1$ and $J_2$ between nearest and next-nearest spins of the Heisenberg $J_1$-$J_2$ spin chain. 
The bare $g_{\rm bs}$ is known to change sign at the critical $J_{2,c} \approx 0.241 J_1$ and is described by $g_{\rm bs} = c (J_{2,c} - J_2)$, with $c > 0$, in the vicinity of the critical point \cite{Eggert1996}. This feature allows one realize the limit of non-interacting spinons (within the low-energy effective theory approximation) by tuning the spin chain to the critical $J_2 = J_{2,c}$ point, and was exploited successfully in Ref.\  \onlinecite{Keselman2020}.

For the chain with DM interaction, Eq.\ \eqref{2.1}, this argument requires modifications beyond the addition of the $J_2$ interaction $J_2 \hat {\mathbf{S}}_n\cdot\hat{\mathbf{S}} _{n+2}$ to the right-hand-side of \eqref{2.1}. In fact, one needs to simulateneously add the DM interaction $D_2$ between the next-nearest spins, that is $\mathbf D_2\cdot \hat{\bf S}_n \times \hat {\bf S}_{n+2}$. The reason for this term is the need to compensate for the generation of the DM-like terms from the $J_2$-part of the Hamiltonian under the unitary rotation \eqref{2.2}. It is straightforward to show that the modified Hamiltonian 
\bea
\hat H_{1+2} 
&=&\sum_n \, J\hat {\mathbf{S}}_n\cdot\hat{\mathbf{S}} _{n+1} + J_2 \hat {\mathbf{S}}_n\cdot\hat{\mathbf{S}} _{n+2} \nonumber\\
&&- D \hat{z} \cdot \hat{\bf S}_n \times \hat {\bf S}_{n+1} - D_2 \hat{z} \cdot \hat{\bf S}_n \times \hat {\bf S}_{n+2} 
\label{ap102}
\eea
transforms under the rotation \eqref{2.2}, with $k_{\rm dm}$ given by \eqref{2.3}, into 
\bea
&&\hat{\tilde{H}}_{1+2}
=\sum_n \, J \sqrt{1+d_1^2} \frac{1}{2}(\hat{\tilde{S}}^+_{n}\hat{\tilde{ S}}^-_{n+1}+ \hat{\tilde{S}}^-_{n}\hat{\tilde{S}}^+_{n+1})
+ J \hat{\tilde{S}}_{n}^z\hat{\tilde{S}}_{n+1}^z \nonumber\\
&&+ J_2 \frac{1+d_1^2}{1-d_1^2} \frac{1}{2}(\hat{\tilde{S}}^+_{n}\hat{\tilde{ S}}^-_{n+2}+\hat{\tilde{S}}^-_{n}\hat{\tilde{S}}^+_{n+2})
+J_2\hat{\tilde{S}}_{n}^z\hat{\tilde{S}}_{n+2}^z ,
\label{ap103}
\eea
{\em provided} that $D_2$ is chosen to be 
\be
D_2 = J_2 \frac{2 d_1}{1-d_1^2} .
\label{ap104}
\ee
Here we abbreviated $d_1 = D/J$. 
Eq.\ \eqref{ap103} is the generalization of \eqref{2.4} to the case of the interaction between both nearest and next-nearest neighbors. 
The effective anisotropy parameters for the nearest spins are $\Delta \approx 1 - d_1^2/2$ (the same as for \eqref{2.4}) and $\Delta_2 \approx 1 - 2 d_1^2$ for the next-nearest ones. Provided that $d_1^2 \ll 1$, which is well satisfied in all considered cases, the Hamiltonian \eqref{ap103} is approximated very well by that of the simple $J_1$-$J_2$ model. 

Correspondingly, $\hat H_{1+2}$ \eqref{ap102} with $D_2$ given by \eqref{ap104} represents the lattice version of the non-interacting spinon limit when $J_2$ is tuned to the vicinity of $J_{2,c}$.  This is the lattice Hamiltonian used in our numerical simulations reported in Fig.\ \ref{fig:dmrg_perp}(c) and Fig.~\ref{fig:dmrg_arbDH}(b). Note that for $D=0.3 J$ used in that calculation, the required value of $D_2 = 0.66 J_{2,c} \approx 0.16 J$ is not particularly small. Still, the absence of any visible splitting between $\omega_0$ and $\omega_1$ branches in  Fig.~\ref{fig:dmrg_perp}(c), as well as an excellent linearity of the obtained spectra near $k=0$, confirm the validity of the described procedure.

\bibliography{ref}
\end{document}